\documentclass[10pt,journal,compsoc]{IEEEtran}

\usepackage{algorithm} 
\usepackage{algorithmic}
\usepackage{booktabs}
\usepackage{clrscode3e}
\usepackage{hyperref}
\usepackage{multirow}
\usepackage{graphicx}
\usepackage{subfigure}
\usepackage{textcomp}
\usepackage{xcolor}
\usepackage{acronym} %for ac command
\usepackage{amsfonts}
\usepackage{amssymb}
\usepackage{amsthm,amsmath}
\usepackage{mathrsfs}
\usepackage{indentfirst}
\usepackage{cases}

\ifCLASSOPTIONcompsoc
  \usepackage[nocompress]{cite}
\else
  \usepackage{cite}
\fi
\ifCLASSINFOpdf
\else
\fi
\acrodef{RL}{Reinforcement Learning}
\acrodef{SCSR}{Shared-account Cross-domain Sequential Recommendation}
\acrodef{SR}{Sequential Recommendation}

\acrodef{RL-ISN}{\textbf{R}einforcement \textbf{L}earning-enhanced \textbf{I}nformation \textbf{S}haring \textbf{N}etwork}
\acrodef{ISN}{Information Sharing Network}
\acrodef{CR}{Cross-domain Recommendation}
\acrodef{CSR}{Cross-domain Sequential Recommendation}
\acrodef{GR}{Group Recommendation}
\acrodef{MDP}{Markov Decision Process}
\acrodef{RL-DF}{Reinforcement learning-enhanced Domain Filter}
\acrodef{RNN}{Recurrent Neural Network}
\acrodef{GRU}{Gated Recurrent Unit}
\acrodef{MLP}{ Multi-Layer Perceptron}
\acrodef{SAM}{Shared-Account Modeling}
\acrodef{DA-GCN}{Domain-Aware Graph Convolutional Network}
\acrodef{BCR}{Basic Cross-domain Recommender}
\acrodef{UIN}{User Identification Network}
\acrodef{GNN}{Graph Neural Network}

\begin{document}
% paper title
\title{Reinforcement Learning-enhanced Shared-account Cross-domain Sequential Recommendation}
%
%
% author names and IEEE memberships
% note positions of commas and nonbreaking spaces ( ~ ) LaTeX will not break
% a structure at a ~ so this keeps an author's name from being broken across
% two lines.
% use \thanks{} to gain access to the first footnote area
% a separate \thanks must be used for each paragraph as LaTeX2e's \thanks
% was not built to handle multiple paragraphs
%
%
%\IEEEcompsocitemizethanks is a special \thanks that produces the bulleted
% lists the Computer Society journals use for "first footnote" author
% affiliations. Use \IEEEcompsocthanksitem which works much like \item
% for each affiliation group. When not in compsoc mode,
% \IEEEcompsocitemizethanks becomes like \thanks and
% \IEEEcompsocthanksitem becomes a line break with idention. This
% facilitates dual compilation, although admittedly the differences in the
% desired content of \author between the different types of papers makes a
% one-size-fits-all approach a daunting prospect. For instance, compsoc 
% journal papers have the author affiliations above the "Manuscript
% received ..."  text while in non-compsoc journals this is reversed. Sigh.

\author{Lei~Guo,
        Jinyu~Zhang,
        Tong~Chen,
        Xinhua~Wang and
        Hongzhi~Yin, ~\IEEEmembership{Senior Member,~IEEE}
        % <-this % stops a space
\IEEEcompsocitemizethanks{\IEEEcompsocthanksitem L. Guo, and J. Zhang are with the School of information science and Engineering, Shandong Normal University, Jinan 250014, China.\protect\\
E-mail: leiguo.cs@gmail.com, jinyuz1996@outlook.com
\IEEEcompsocthanksitem T. Chen is with the School of Information Technology \& Electric Engineering, The University of Queensland, St Lucia, QLD 4072, Australia.\protect\\
E-mail: tong.chen@uq.edu.au
\IEEEcompsocthanksitem X. Wang is with the School of information science and Engineering, Shandong Normal University, Jinan 250014, China.\protect\\
E-mail: wangxinhua@sdnu.edu.cn
\IEEEcompsocthanksitem H. Yin is with the School of Information Technology \& Electric Engineering, The University of Queensland, St Lucia, QLD 4072, Australia.\protect\\
E-mail: h.yin1@uq.edu.au

% \IEEEcompsocthanksitem J. Doe and J. Doe are with Anonymous University.
}% <-this % stops an unwanted space
\thanks{Manuscript received XX XX, 2022; revised XX XX, 2022.\\
(Corresponding author: Hongzhi Yin.)}}

% note the % following the last \IEEEmembership and also \thanks - 
% these prevent an unwanted space from occurring between the last author name
% and the end of the author line. i.e., if you had this:
% 
% \author{....lastname \thanks{...} \thanks{...} }
%                     ^------------^------------^----Do not want these spaces!
%
% a space would be appended to the last name and could cause every name on that
% line to be shifted left slightly. This is one of those "LaTeX things". For
% instance, "\textbf{A} \textbf{B}" will typeset as "A B" not "AB". To get
% "AB" then you have to do: "\textbf{A}\textbf{B}"
% \thanks is no different in this regard, so shield the last } of each \thanks
% that ends a line with a % and do not let a space in before the next \thanks.
% Spaces after \IEEEmembership other than the last one are OK (and needed) as
% you are supposed to have spaces between the names. For what it is worth,
% this is a minor point as most people would not even notice if the said evil
% space somehow managed to creep in.

% The paper headers

\markboth{IEEE TRANSACTIONS ON KNOWLEDGE AND DATA ENGINEERING,~Vol.~XX, No.~X, X~2022}
% 这里定义第second header
{Shell \MakeLowercase{\textit{et al.}}: Bare Demo of IEEEtran.cls for Computer Society Journals}
% The only time the second header will appear is for the odd numbered pages
% after the title page when using the twoside option.
% 
% *** Note that you probably will NOT want to include the author's ***
% *** name in the headers of peer review papers.                   ***
% You can use \ifCLASSOPTIONpeerreview for conditional compilation here if
% you desire.

% The publisher's ID mark at the bottom of the page is less important with
% Computer Society journal papers as those publications place the marks
% outside of the main text columns and, therefore, unlike regular IEEE
% journals, the available text space is not reduced by their presence.
% If you want to put a publisher's ID mark on the page you can do it like
% this:
%\IEEEpubid{0000--0000/00\$00.00~\copyright~2015 IEEE}
% or like this to get the Computer Society new two part style.
%\IEEEpubid{\makebox[\columnwidth]{\hfill 0000--0000/00/\$00.00~\copyright~2015 IEEE}%
%\hspace{\columnsep}\makebox[\columnwidth]{Published by the IEEE Computer Society\hfill}}
% Remember, if you use this you must call \IEEEpubidadjcol in the second
% column for its text to clear the IEEEpubid mark (Computer Society jorunal
% papers don't need this extra clearance.)

% use for special paper notices
%\IEEEspecialpapernotice{(Invited Paper)}

% for Computer Society papers, we must declare the abstract and index terms
% PRIOR to the title within the \IEEEtitleabstractindextext IEEEtran
% command as these need to go into the title area created by \maketitle.
% As a general rule, do not put math, special symbols or citations
% in the abstract or keywords.
\IEEEtitleabstractindextext{%
\begin{abstract}
\acf{SCSR} is an emerging yet challenging task that simultaneously considers the shared-account and cross-domain characteristics in the sequential recommendation. Existing works on \ac{SCSR} are mainly based on \acf{RNN} and \acf{GNN} but they ignore the fact that although multiple users share a single account, it is mainly occupied by one user at a time. This observation motivates us to learn a more accurate user-specific account representation by attentively focusing on its recent behaviors. Furthermore, though existing works endow lower weights to irrelevant interactions, they may still dilute the domain information and impede the cross-domain recommendation.
To address the above issues, we propose a reinforcement learning-based solution, namely RL-ISN, which consists of a basic cross-domain recommender and a reinforcement learning-based domain filter. Specifically, to model the account representation in the shared-account scenario, the basic recommender first clusters users' mixed behaviors as latent users, and then leverages an attention model over them to conduct user identification. To reduce the impact of irrelevant domain information, we formulate the domain filter as a hierarchical reinforcement learning task, where a high-level task is utilized to decide whether to revise the whole transferred sequence or not, and if it does, a low-level task is further performed to determine whether to remove each interaction within it or not.
To evaluate the performance of our solution, we conduct extensive experiments on two real-world datasets, and the experimental results demonstrate the superiority of our RL-ISN method compared with the state-of-the-art recommendation methods.
\end{abstract}

% Note that keywords are not normally used for peerreview papers.
}

% make the title area
\maketitle

% To allow for easy dual compilation without having to reenter the
% abstract/keywords data, the \IEEEtitleabstractindextext text will
% not be used in maketitle, but will appear (i.e., to be "transported")
% here as \IEEEdisplaynontitleabstractindextext when the compsoc 
% or transmag modes are not selected <OR> if conference mode is selected 
% - because all conference papers position the abstract like regular
% papers do.
\IEEEdisplaynontitleabstractindextext
% \IEEEdisplaynontitleabstractindextext has no effect when using
% compsoc or transmag under a non-conference mode.

% For peer review papers, you can put extra information on the cover
% page as needed:
% \ifCLASSOPTIONpeerreview
% \begin{center} \bfseries EDICS Category: 3-BBND \end{center}
% \fi
%
% For peerreview papers, this IEEEtran command inserts a page break and
% creates the second title. It will be ignored for other modes.
\IEEEpeerreviewmaketitle

% \IEEEraisesectionheading{\section{Introduction}\label{sec:introduction}}
% Computer Society journal (but not conference!) papers do something unusual
% with the very first section heading (almost always called "Introduction").
% They place it ABOVE the main text! IEEEtran.cls does not automatically do
% this for you, but you can achieve this effect with the provided
% \IEEEraisesectionheading{} command. Note the need to keep any \label that
% is to refer to the section immediately after \section in the above as
% \IEEEraisesectionheading puts \section within a raised box.

% The very first letter is a 2 line initial drop letter followed
% by the rest of the first word in caps (small caps for compsoc).
% 
% form to use if the first word consists of a single letter:
% \IEEEPARstart{A}{demo} file is ....
% 
% form to use if you need the single drop letter followed by
% normal text (unknown if ever used by the IEEE):
% \IEEEPARstart{A}{}demo file is ....
% 
% Some journals put the first two words in caps:
% \IEEEPARstart{T}{his demo} file is ....
% 
% Here we have the typical use of a "T" for an initial drop letter
% and "HIS" in caps to complete the first word.
% \IEEEPARstart{T}{his} demo file is intended to serve as a ``starter file''
% for IEEE Computer Society journal papers produced under \LaTeX\ using
% IEEEtran.cls version 1.8b and later.
% % You must have at least 2 lines in the paragraph with the drop letter
% % (should never be an issue)
% I wish you the best of success.

% \hfill mds
 
% \hfill August 26, 2015

% \subsection{Subsection Heading Here}
% Subsection text here.
\IEEEraisesectionheading{\section{Introduction}\label{sec:introduction}}

\IEEEPARstart{I}{n} many online systems (such as E-commerce and social media sites), user interactions are organized into sequences, i.e., organizing user behaviors in chronological order, making \ac{SR} a hot research topic. Moreover, as users tend to sign up for different platforms to access domain-specific services, e.g., news subscription and video watching, \ac{CSR} that aims at making the next item recommendation via leveraging users' historical sequential behaviors from multiple domains is gaining immense attention.
In this work, we study \ac{CSR} in an emerging yet challenging context, \ac{SCSR}, in which multiple users share a single account, and their interactions are mixed in multiple domains. 
% We consider the shared account characteristic because it is a common phenomenon that people tend to share accounts with others in practice. For example, members of a family tend to watch videos and shop online through shared accounts. The mixture of users' behaviors and diverse interests not only makes generating accurate recommendations more difficult, but also amplifies the noise in the interaction data and impedes the cross-domain recommendation.
We consider the shared-account characteristic because account sharing has become pervasive in many Internet applications and services. For example, friends or family members tend to share one account in watching the smart TV or speaking to the smart speaker, e.g., Tmall Genie or Google Home Hub, and share one premium account to enjoy online streaming services provided by YouTube or Netflix.
Another example is that all persons in a household tend to share one music-streaming account and one online-shopping account.
Moreover, each member behind a shared-account may have different interests and recommendation demands, and it is impractical to view a shared-account as a single user, which further motivates us to design recommenders for shared-accounts.
However, \ac{SCSR} is a challenging task since the mixture of users' behaviors and diverse interests not only makes generating accurate recommendations more difficult, but also amplifies the noise in the interaction data and impedes the cross-domain recommendation.
Note that, sharing 
% an account, a kind of user privacy, 
is an active behavior among close friends or family members, and our task is to make recommendations for these existing shared-accounts, rather than linking two or multiple accounts that potentially belong to the same person, i.e., leaking users' privacy. In fact, shared-accounts can resist malicious attacks (for both attribute inference and membership inference attacks), as it mixes multiple users' behaviors.
% as they have mixed user behaviors and know nothing about the membership information.

Recently, several studies have been focused on conducting cross-domain or shared-account recommendations, but they rarely address the \ac{SCSR} setting that considers both characteristics.
Prior works on cross-domain recommendations mainly focus on aggregating information from both domains~\cite{zhuang_cross_2010,cao_transfer_2010,abel2013cross,Zhu2021crossdomain} or transferring knowledge from the source domain to the target domain~\cite{hu2018conet, Fan2021crossdomain, Salah2021crossdomain}. The main assumption behind \ac{CSR} is that users might have similar interests in different domains, and the user behaviors in one domain have the potential to improve recommendations in another domain. However, as users have diverse preferences, and some of them are domain-specific, the effect of the behaviors that reflect the users' interests in the target domain might be diluted by many irrelevant interactions from the source domain. More importantly, as there is no direct supervision, distinguishing the noisy information transferred from other domains is still a challenging task. 
In addition, existing methods cannot be directly applied to \ac{SCSR}, since the important shared-account characteristic is largely ignored, and they can only make recommendations for single users.
In previous studies on the shared-account recommendation~\cite{zhao2016passenger,jiang2018identifying,wen_miss_2021}, a common solution is to extract latent representations from high-dimensional feature spaces that describe users' relationships under the same account. However, most of them are focused on a single domain, and hence are inapplicable to \ac{SCSR}.

One recent study that focuses on solving \ac{SCSR} is $\pi$-net~\cite{ma2019pi}, which is a parallel sequential recommender with \ac{RNN} as its basic sequence encoder and a gated unit as its cross-domain transfer controller to share information between two domains. Another prior work that addresses \ac{SCSR} is DA-GCN~\cite{guo_DAGCN_2021}, which proposes a graph-based solution with a cross-domain sequential graph to explicitly link accounts and items from two domains. In their work, a domain-aware graph convolutional network is devised to learn expressive representations for items and account-sharing users.
Though the above methods have improved the recommendation performance by exploring the cross-domain and shared-account characteristics, the \ac{SCSR} task is still largely unexplored due to the following reasons. First, as users usually have diverse interests, their interactions in one domain may not reflect their interests in another domain. Instead, their interests might be diluted by irrelevant domain information. Despite existing models assign relatively high attention coefficients to the major contributing interactions, the effect of the domain information is still discounted by the irrelevant ones.
Second, existing methods mainly divide the learned account representation evenly in a high-dimensional space to obtain the representation of each latent user, since the unavailability of the user identity information. We argue that the users' representations can be possibly identified, as an account is mainly used by one user at a time, and current interaction behaviors of an account expose a user's real interest.
% cross-target-domain

To address the above issues, we propose a \ac{RL}-based solution, namely \ac{RL-ISN}, for \ac{SCSR}. Our solution consists of two modules: a \ac{BCR} and a \ac{RL-DF}. Specifically, to model accounts' representations under the shared-account context, \ac{BCR} first maps the mixed user behaviors into a high-dimensional feature space via a fully connected layer, where a hidden node is deemed as a latent user. Then, we learn the account representation by attentively aggregating the latent users within it, and then enhance target domain recommendation by transferring information from the source domain. 
Moreover, since the irrelevant domain information may hurt the cross-domain recommendation in particular without the explicit/supervised information about the domain knowledge, we formalize the knowledge transfer as a hierarchical reinforcement learning task (i.e., the \ac{RL-DF} module). That is, a high-level task is performed to decide whether to revise the whole transferred sequence or not, and if it does, a low-level task is performed to determine whether to keep each interaction within the sequence.
% We conduct extensive experiments on two real-world datasets, and the experimental results demonstrate the advantage of our \ac{RL-ISN} method compared with the state-of-the-art recommendation methods.

The main contributions of this work are listed as follows.

\begin{itemize}
    \item We propose a novel \ac{RL-ISN}, which consists of a basic cross-domain recommender and a domain filter, to encode and share the domain information.
    \item We devise an attention-based cross-domain sequence encoder as the basic recommender to model the information from multiple domains. To consider the shared-account characteristic, we further develop a user identification network by clustering and identifying the latent users sharing the same account.
    \item We develop a \ac{RL}-based domain filter to retain only the  interactions that are helpful for cross-domain recommendations via considering the rewards brought by the transferred domain knowledge.
    \item We conduct extensive experiments on two real-world datasets to demonstrate the advantage of our proposed \ac{RL-ISN} method compared with several state-of-the-art recommendation methods. 
\end{itemize}
% \vspace{-0.5cm}
\section{Related Work}
% \noindent This section considers four types of recommendation methods, i.e., sequential recommendation, shared-account recommendation, cross-domain recommendation and reinforcement learning-based recommendation, as our related works.

\subsection{Sequential Recommendation}
\noindent The task of \ac{SR} is to predict the next item that a user tends to interact with given her historical interactions. Traditional studies on \ac{SR} are mainly Markov chain-based methods, which treat the recommendation generation as predicting the next action in a \ac{MDP}~\cite{JMLR_shani05a, chen_playlist_2012}. For example, Shani et al.~\cite{JMLR_shani05a} propose a \ac{MDP}-based recommender by taking into account both the long-term effects and the expected
value of each recommendation.
Chen et al.~\cite{chen_playlist_2012} formulate the playlist generation problem as a regularized maximum-likelihood embedding of Markov chains in Euclidean space and solve it by devising a logistic Markov embedding algorithm.
However, the above methods have limited ability in modeling sequences, since their state space becomes unmanageable when considering the whole sequence.
To enhance the capability of capturing the high-dimensional expression of sequences, recent works are mainly focused on devising deep neural network-based methods~\cite{hidasi2016srnn,quadrana2017hrnn,li2017neural,he2018nais,Xie2021sequential,QiuHLY20, ChenYCNPL19}. For example, Hidasi et al.~\cite{hidasi2016srnn} apply \ac{RNN} to session-based sequential recommendation, and achieve significant improvement over traditional methods. 
Hsu et al.~\cite{Hsu2021sequential} propose a temporal attentive graph neural network for \ac{SR}, and can achieve the state-of-the-art performance under conventional, inductive, and transferable settings.
% Xie et al. ~\cite{Xie2021sequential} pay their attention to the rich heterogeneous information such as multi-typed interactions and item attributes. They integrate these rich information by introducing  dynamic heterogeneous information networks.
However, these sequential recommenders are mainly focused on capturing the sequential characteristics for single users, and their performance on the shared-account and cross-domain scenarios are largely unexplored.

\subsection{Shared-account Recommendation}
\noindent Previous studies on the shared-account recommendation are usually divided into two steps, that is, first perform user identification, and then make recommendations~\cite{zhao2016passenger,jiang2018identifying,wang2014user}. For example, Zhang et al.~\cite{zhang2012guess} first study user identification as a subspace clustering problem, and then show the possible improvement it brings to personalized recommendation. Jiang et al.~\cite{jiang2018identifying} propose a session-based heterogeneous graph to embed different users under an account, where the items and their available metadata are both considered. 
% Nery et al.~\cite{Nery2021shareaccount} present a new method to firstly identify online sessions on a platform and afterward, to cluster these sessions to identify the profiles of the latent users behind the shared accounts.
Wang et al.~\cite{wang2014user} model users by utilizing their consumption logs with the assumption that different users are active in different times periods, and further leverage a standard KNN to make recommendations.
Similar to~\cite{wang2014user}, Yang et al.~\cite{yang2015adaptive} judge whether the split historical behaviors belong to the same user or not by analyzing their similarity.
However, the above methods can only conduct user identification and recommendation in two separate processes, and ignore the benefit brought by the end-to-end learning. Though Wen et al.~\cite{wen_miss_2021} recently develop a unified multi-user identification network based on a self-attentive method, they ignore the cross-domain information, which may further improve the shared-account recommendation.

\subsection{Cross-domain Recommendation}
\noindent \acf{CR} that aims at improving recommendations by concerning data from multiple domains has been proven helpful for cold-start and sparsity issues~\cite{abel2013cross,Zhu2021crossdomain}.
Traditional methods for \ac{CR} can be categorized into knowledge aggregation-based methods and knowledge transferring-based methods. In aggregation-based methods, the efforts are making at designing aggregating functions that can take into account both domains~\cite{zhuang_cross_2010,cao_transfer_2010}. In transferring-based methods, the studies are mainly focused on sharing or transferring domain information between two domains~\cite{hu2018conet, Salah2021crossdomain}. 
However, these two kinds of methods are all shallow methods, which hinders them to learn high-level abstractions from different domains.
In light of this, deep neural network-based methods have gradually attracted the attention of researchers~\cite{chengzhao2019ppgn, ChenYS0GM20,liu2020transfer,lian2017cccfnet,Zhu2021crossdomain, Li2021crossdomain}. 
For example, Liu et al.~\cite{LiuSCZ21} propose a distribution alignment-based method to make \ac{CR} for the cold-start items in the target domain. But their method needs items' auxiliary information as input and can only make recommendations for a single target.
In light of the limitations of single-target \ac{CR}, Zhu et al.~\cite{ZhuWCLZ20} first propose a graphical and attentional framework for dual-target \ac{CR} to improve the recommendation accuracy on both domains. Then, they extend~\cite{ZhuWCLZ20} in~\cite{zhu2021unified}, and further devise a unified framework for multi-types of \ac{CSR}s via a personalized training strategy.
% \textcolor{red}{For example, 
% Liu et al.~\cite{LiuSCZ21} focus on studying the cross-domain cold-start recommendation 
% Cross-Domain Cold-Start Recommendation task (CDCSR), which aims at leveraging the information from a source domain, where items are `warm', to improve the recommendation performance of a target domain, where items are `cold'.
% CDCSR is different from our SCSR task, since CDCSR assumes both domains have auxiliary representations such as item profiles or descriptions, and aims at making single recommendations for the items in the target domain, while SCSR only needs the users' sequential behaviors as input, and makes dual recommendations for the shared-accounts in both domains.
% Zhu et al.~\cite{ZhuWCLZ20} first propose a graphical and attentional framework for dual-target \ac{CR} to improve the recommendation accuracy on both domains. Then, they extend \cite{ZhuWCLZ20} in \cite{zhu2021unified}, and further devise a unified framework for multi-types of \ac{CSR}s via a personalized training strategy.
% ~\cite{ZhuWCLZ20, zhu2021unified} propose a graphical framework for dual-target Cross-Domain Recommendation (DTCDR), they utilize an element-wise attention network to combine the information from different domains.
% Liu et al.~\cite{LiuSCZ21} utilizes both rating and auxiliary information to learning the user/item representation across domain for the Cross-Domain Cold-Start Recommendation (CDCSR) task. }
However, the above methods all ignore the fact that irrelevant domain information, even if they are given lower attention weights, may still dilute the domain knowledge. Moreover, these methods do not consider the shared-account scenario, which may further amplify the noisy interactions and prevent the cross-domain recommendation.

\subsection{Reinforcement Learning-based Recommendation}
\noindent Recently, reinforcement learning-based technique has been widely applied to solve the recommendation issues, such as the noisy items in sequential recommendation~\cite{Zhang2019HierarchicalRL}, dialog-based interactive recommendation~\cite{wu2021reinforcement}, news recommendation ~\cite{zheng2018reinforcement} and social recommendation~\cite{Zhang2021reinforcement, Sun2021reinforcement}. For example,  Wu et al. ~\cite{wu2021reinforcement} develop a dialog-based recommendation model by leveraging an estimator to track and estimate users’ preferences, and a generator to match the estimated preferences with the candidate items to make recommendations. Zheng et al.~\cite{zheng2018reinforcement} introduce a deep reinforcement learning framework for  online personalized news recommendation by utilizing deep Q-learning to model both current and future reward.
% Zhang et al. ~\cite{Zhang2021reinforcement} suppose to help the electric vehicle drivers to find the proper spots for charging by using  proposed multi-agent reinforcement learning network, which using the centralized attentive mechanism to coordinate the recommendation between multiple agents.  
Zhang et al.~\cite{Zhang2019HierarchicalRL} propose a hierarchical reinforcement learning-based algorithm for course recommendation by tackling the noises existed in users' historical courses, where a user profile reviser is devised to remove the courses that cannot bring more rewards to recommender.
% These methods are usually implemented by one or multiple agents to execute the semi Markov decision process (SMDP) ~\cite{wang2018reinforcement,williams1992simple,wu2021reinforcement,Fan2021crossdomain}.
% And most of the algorithms rely on a delayed reward mechanism to update the agent's policy for the next decision. ~\cite{Zhang2019HierarchicalRL,Zhang2021reinforcement,Sun2021reinforcement} For instance, Wu et al. ~\cite{wu2021reinforcement} propose an Estimator-Generator-Evaluator model, with Q-learning for partially observable Markov decision process (POMDP), to effectively incorporate the users’ preferences over time. Zhang et al. ~\cite{Zhang2021reinforcement} develop a framework with multiple agents to improve the charging experience from various aspects over a long-term horizon for the electric vehicles. With the application of deep Q-learning, Zheng et al. ~\cite{zheng2018reinforcement} propose a novel reinforcement learning framework for news recommendation. Recently, some researchers attempt to use the hierarchical reinforcement learning method to decompose the complex tasks into multiple small tasks to reduce the overall complexity of recommendation task \cite{sutton2000policy,williams1992simple}. For example, Zhang et al. ~\cite{Zhang2019HierarchicalRL} propose a hierarchical reinforcement learning algorithm to revise the user profiles for addressing the challenge that user profiles were diluted by diverse historical behaviors. 
However, although the existing reinforcement learning-based methods have achieved great success in many recommendation tasks, none of them have been focused on solving the cross-domain transfer issue and the shared-account characteristic. Due to the missing of the direct supervision for the domain information, we cast the process of domain transfer as a \ac{MDP}, and further improve the cross-domain recommendation by a hierarchical reinforcement learning network.

\subsection{Shared-account Cross-domain Sequential Recommendation}
\noindent To simultaneously explore the shared-account and cross-domain characteristics, recent studies have given more attention to the \ac{SCSR} task. Ma et al.~\cite{ma2019pi} are the first to study \ac{SCSR}, and propose a parallel information sharing network, namely $\pi$-net, which uses a \ac{RNN}-based unit to learn user-specific representations for $k$ latent users, and a gating mechanism to filter out domain information. Ren et al.~\cite{ren2019net} further improves $\pi$-net by first splitting the role-specific representations at each timestamp and then joining them to get cross-domain representations.
Guo et al.~\cite{guo_DAGCN_2021} develop a graph-based solution to enhance the expressive ability of sequential patterns via capturing the structure of cross-domain information. However, none of the above methods take into account the fact that an account is mainly occupied by one user at a time, and the current user can be possibly identified by her recent behaviors. Moreover, existing methods that endow lower weights to irrelevant interactions may also dilute the effect of the domain information and hinder the cross-domain recommendation.
\begin{figure*}
    \centering
    \includegraphics[width=12.5cm]{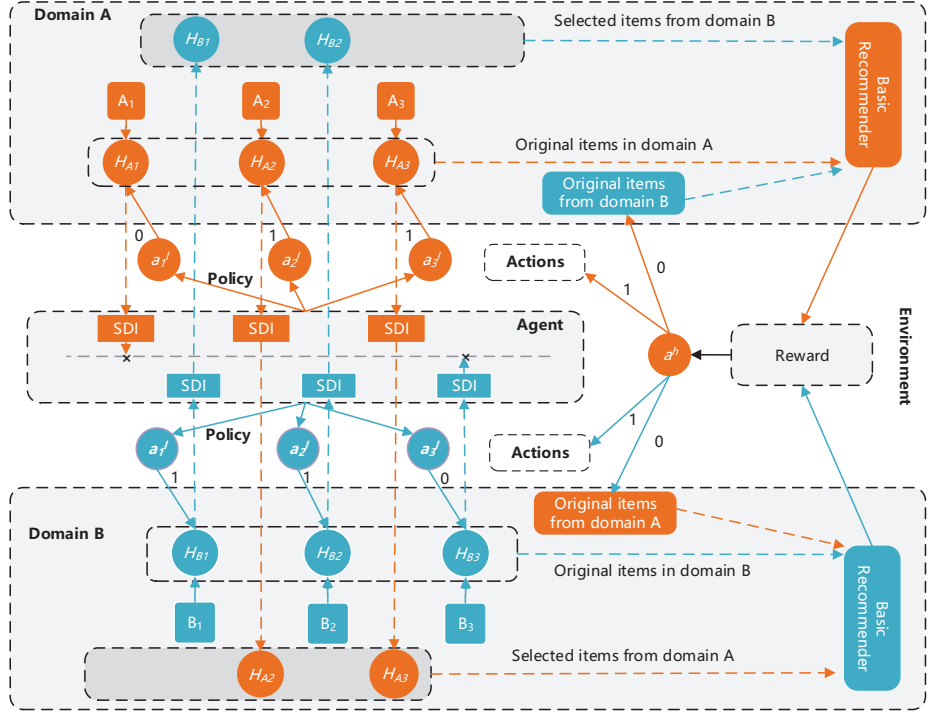}
    \caption{The System Architecture of \ac{RL-ISN}, where SDI stands for the Shared Domain Information.}
    \label{fig:framework}
\end{figure*}

\section{Methodologies\label{sec:methodologies}}
% \noindent In this section, we first formulate the \ac{SCSR} problem, and then describe our reinforcement learning-enhanced cross-domain recommender.

\subsection{Preliminaries}
% We use bold capital letters (e.g., $\boldsymbol{X}$) and bold lowercase letters (e.g., $\boldsymbol{x}$) to represent matrices and vectors, respectively. We employ squiggle letters (e.g., $\mathcal{X}$) to denote sets, and none-bold lowercase or capital letters (e.g., $x$ or $X$) to denote scalars. All vectors are in column forms if not clarified.

\noindent The task of \ac{SCSR} is to predict the next items that a shared-account will consume in different domains. For simplicity, this work only considers two domains but the ideas can be easily generalized to multiple domains. \ac{SCSR} is different from traditional \ac{SR} task in two aspects. First, in \ac{SCSR}, the sequential behaviors are generated by multiple users, while in the traditional \ac{SR} they are usually generated by a single user. Second, \ac{SCSR} considers the information from multiple domains for a particular recommendation, while \ac{SR} only considers the information in a single domain.
% Note that, \ac{SCSR} is also different from \ac{GR}~\cite{yin2019social,guo2020group}, since \ac{GR} can make recommendations by leveraging the group members' information and preferences, while \ac{SCSR} does not have such membership information. 
Here, we give a formulation of the \ac{SCSR} task.

Suppose there are $C$ shared-accounts $\mathcal{G}=\{ G_1, G_2, ..., G_c, ..., G_C \}$, $M$ items $\mathcal{V}_A= \{ A_1, A_2, ..., A_m, ..., A_M \}$ in domain A and $N$ items $\mathcal{V}_B=\{B_1, B_2, ..., B_n, ..., B_N\}$ in domain B. 
% Each account is shared by $L$ users $(u_1, u_2, ..., u_l, ..., u_L)$. 
We represent the behavior sequence generated by account $G_c$ as $\mathcal{S}^o_c=\{A_1, B_1, ..., A_m, ..., B_n, ...\}$,  where $A_m\in \mathcal{V}_A$ is a consumed item in domain A, and $B_n \in \mathcal{V}_B$ is a consumed item in domain B. Given $\mathcal{S}^o_c$, \ac{SCSR} aims to recommend a ranked list of items that account $G_c$ would like to consume next in domain A or B, which can be formally defined as:

\textbf{Input:} A shared-account $G_c$, the item set $\mathcal{V}_A$ and $\mathcal{V}_B$, the mixed historical behavior sequence $\mathcal{S}^o_c$ generated by $G_c$ in domain A and B.

\textbf{Output:} 1) The probability $P(A_m|\mathcal{S}^o_c)$ of recommending item $A_m$ as the next item to be consumed in domain A, and
2) the probability $P(B_n|\mathcal{S}^o_c)$ of recommending item $B_n$ as the next item to be consumed in domain B.

\subsection{Overview of RL-ISN}
\noindent The purpose of \ac{RL-ISN} is to learn a cross-domain recommender that can transfer the refined domain knowledge under the shared-account scenario from domain A to B to improve the recommendation performance in B, and vice versa.
Since there is no prior knowledge about user identities in an account, directly modeling users' mixed sequential behaviors is not feasible. Hence, we leverage a fully connected layer to first conduct behavior clustering, and then identify users via the current target and an attention network.
% \textcolor{red}{Besides, to consider the temporality of users' mixed historical behaviors, we inject the positional encoding to the input sequences to match items and their original sequential orders.} 
Moreover, as the recommendation process may be disturbed by irrelevant domain information, we need to learn an agent to filter out the irrelevant interactions from the transferred domain information, and then make recommendations based on the refined transferred knowledge.
The key challenge here is how to determine which transferred interactions are the noises without direct supervision signal. To deal with this issue, we formalize the process of transferring knowledge across domains as a hierarchical \ac{MDP} and resort to the hierarchical reinforcement learning technique, where a high-level and a low-level task are performed to fine-tune the transferred domain information.

\begin{figure*}
    \centering
    \includegraphics[width=14.5cm]{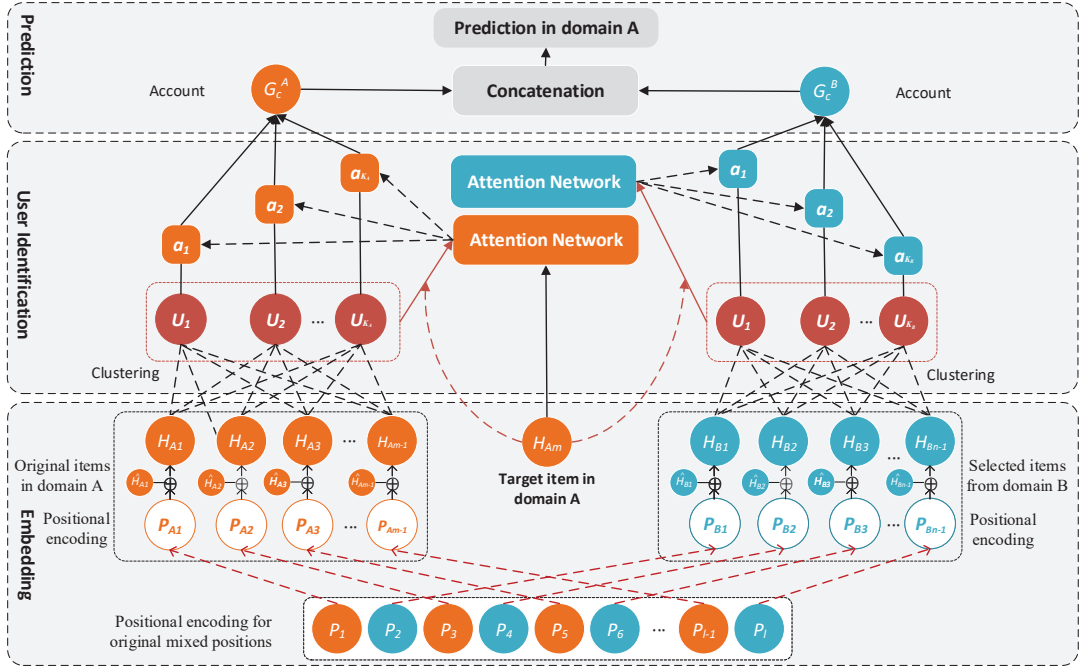}
    \caption{Overview of the basic cross-domain recommender in domain A. A similar basic recommender in domain B can also be easily achieved.}
    \label{fig:basic_recommender}
\end{figure*}

Fig.~\ref{fig:framework} shows the architecture of our \ac{RL-ISN}, which consists of two components: a \acf{BCR} and a \acf{RL-DF}.
1) To encode users' mixed behaviors, \ac{BCR} (Fig.~\ref{fig:basic_recommender} shows an overview of \ac{BCR} in domain A) first clusters the interactions as latent users and then identifies the currently active ones by their recent intents (i.e., the last interaction or the target item). After that, we can achieve the account representation by attentively aggregating the identified latent users in an attention network. 
To make cross-domain recommendations, the information from domain B (or A) is further combined with the original information in domain A (or B) (more details can be seen in Section \ref{sec:basic_recommender}).
% To accelerate the recommendation process, we jointly train the recommendation process in both domains (more details can be seen in Section \ref{sec:basic_recommender}).
2) The \ac{RL-DF} component (or agent) aims to find out a policy that determines which interactions can be transferred across domains, i.e., filtering out the irrelevant user behaviors that may harm the cross-domain recommendation. 
Specifically, we decompose the overall \ac{MDP} process into a high-level task to determine whether to revise the whole transferred sequence or not, and a low-level task to decide whether to remove an interaction within it.
At the end of the revising process, the agent gets both immediate rewards and a delayed long-term reward from the environment, based on which it updates its policy.
As desirable, the performance of \ac{BCR} could be improved by the revised domain information. To simultaneously enhance \ac{BCR} and \ac{RL-DF}, we train the two components in a joint way (more details can be seen in Section \ref{domain_adapter}).

\subsection{Basic Cross-domain Recommender\label{sec:basic_recommender}}
\noindent The key function of our \ac{BCR} module is to accurately characterize an account's preference according to its historical behaviors. However, since each account is shared by multiple users and their behaviors are mixed together, it is not feasible to model them directly.
Moreover, unlike linguistic sequences that are generated in a strictly-ordered way, user behavior may not be in such a strict-chronological order~\cite{xia_self_AAAI_2021}. Strictly modeling the relative orders of items would probably make the recommendation models prone to overfitting.
To overcome the above challenges, we devise a user identification network to model shared-accounts from a multi-user view, and encode accounts' representations by attentively aggregating each latent user within them.

As shown in Fig.~\ref{fig:basic_recommender}, our basic recommender consists of three components, i.e., embedding layer, user identification network, and the prediction component.
In the following, we will detail each of them.

\subsubsection{Embedding Layer}
\noindent For all the items in domains A and B, we respectively create two item embedding matrices $\boldsymbol{M}\in \mathbb{R}^{M\times d}$ and $\boldsymbol{N}\in \mathbb{R}^{N\times d}$, where $d$ is the embedding dimensionality. For each training sequence in A $\{A_1, A_2, ..., A_m, ..., A_{l_a}\}$ or B $\{B_1, B_2, ..., B_n, ..., B_{l_b}\}$, we transform it into a fixed-length sequence according to the maximum sequence length (denoted by $l_a$ or $l_b$) in the current batch. That is, if a sequence length is less than $l_a$ (or $l_b$), we repeatedly add a padding item to the left until its length is $l_a$ (or $l_b$). Then, we retrieve the input items from $\boldsymbol{M}$ (or $\boldsymbol{N}$) as their initial embedding matrix $\boldsymbol{\hat{H}}\in \mathbb{R}^{l_a \times d}$ (or $\boldsymbol{\hat{H}}\in \mathbb{R}^{l_b \times d}$).
We assign a constant zero vector to the padding item. To identify the current user who is using the account, we take the last item (or target item) of each sequence as the supervision signal.

\textbf{Positional Embedding.} To consider items' sequential orders, we first introduce a learnable embedding $\boldsymbol{P}\in \mathbb{R}^{(l_a+l_b)\times d}$~\cite{kang2018self} to encode all the positions of items in the mixed sequences, which is shared by all the items in both domains and invariant to different items in the same position. Then, we inject it into the initial embedding layer (take domain A as an example):
\begin{align}
    \boldsymbol{{H}_A}={
    \left[ \begin{array}{c}
    \boldsymbol{\hat{H}}_{A_1} + \boldsymbol{P}_{A_1} \\
    \boldsymbol{\hat{H}}_{A_2} + \boldsymbol{P}_{A_2} \\
    ...\\
    \boldsymbol{\hat{H}}_{A_m} + \boldsymbol{P}_{A_m} \\
    ...\\
    \boldsymbol{\hat{H}}_{l_a-1} + \boldsymbol{P}_{l_a-1}
    \end{array} 
    \right ]},
\end{align}
where $\boldsymbol{{H}_A}$ is the representation of the given sequence with position embedding, $\boldsymbol{\hat{H}}_{A_m}$ is the initial embedding of item $A_m$, and $\boldsymbol{P}_{A_m}$ is the embedding of the position that item $A_m$ lies in its original sequence.

\subsubsection{User Identification Network}
\noindent As the interactions in a shared-account are generated by multiple users and each user has diverse interests, modeling an account by a single embedding vector is insufficient. To capture the shared-account characteristics, we first cluster the mixed behaviors as latent users, and represent a user via a cluster over all the historical interactions with different aggregation weights.
Moreover, as an account is mainly occupied by one single user at a time, we then represent the account by attentively aggregating these latent users with the target item as the supervision signal.

\textbf{Latent User Representation.} Through the embedding layer, we can obtain the embeddings of all the historical items generated by an account $\{\boldsymbol{H}_{A_1}, \boldsymbol{H}_{A_2}, ..., \boldsymbol{H}_{A_m}, ..., \boldsymbol{H}_{A_{l_a-1}}\}$ in domain A (take domain A as an example). 
% We leverage $\boldsymbol{H}_{A_{t}}$ to  denote the embedding of the target item in A.
% (or $\{\boldsymbol{H}_{B_1}, \boldsymbol{H}_{B_2}, ..., \boldsymbol{H}_{B_n}, ..., \boldsymbol{H}_{B_{l_b-1}}\}$ in domain B).
% We denote the target item in A as $\boldsymbol{H}_{A_{l_a}}$. 
In this work, we adopt a \ac{MLP} network to generate user clusters (assuming there are $K_A$ and $K_B$ latent users in domain A and B, respectively).
Due to there is no supervision signal indicating which user interacts which item, we represent the latent user by considering all the items within this sequence. More specifically, we utilize a fully connected feed-forward network (as shown in Fig.~\ref{fig:basic_recommender}) to map the historical interactions into a high-dimensional user latent feature space to represent each user $u_i$, and denote the mapping function by a linear transformation with a ReLU activation:
\begin{align}
    \boldsymbol{U}_i = \text{max} (0, \sum_{m=1}^{l_a-1} w_{mi} \boldsymbol{H}_{A_m} + b ),
\end{align}
where $\boldsymbol{U}_i$ is the representation of user $u_i$, $w_{mi}$ is the learned weight denoting the importance of the item $A_m$, and $b$ is the bias term.
% Due to there is no supervision signal indicating which user the item belongs to, we represent the latent user by considering all the items within this sequence, and then identify them with the current target item.

To obtain all latent users' representations, we need to map all the interactions within the input sequence $K_A$ (or $K_B$) times. Thus, we extend the weight $w_{mi}$ into a matrix $\boldsymbol{W}^{{K_A} \times (l_a -1)}$, and represent the final matrix of latent users as:
\begin{align}
    \boldsymbol{U} = max (0, \boldsymbol{W} \boldsymbol{H}_c^A + \boldsymbol{b}),
\end{align}
where $\boldsymbol{H}_c^A\in \mathbb{R}^{(l_a-1)\times d}$ denotes the embedding matrix of the whole sequence.

\textbf{Current Account Representation.}
As an account is mainly occupied by one user at a time, its representation should be consistent with that of the user currently using it. In light of this, we take the last item in the current sequence as the supervision signal, and represent the account by attentively aggregating all the latent users under it, that is, an attention mechanism is adopted to estimate the attention weights of latent users in the account representation. 

For the $i$-th latent user $u_i$, we compute how the target item $A_{t}$ is consistent with $u_i$ as:
\begin{align}
    a_{i, t} = f(\boldsymbol{H}_{A_{t}}, \boldsymbol{U}_i),
    % a_c = \text{softmax} (H_{a_t} \boldsymbol{U}_i^T),
\end{align}
where $f$ is an attention function. Inspired by the recent success of modeling the attention weight with neural networks~\cite{guo2020group,li2017neural}, we similarly use a \ac{MLP} to parameterize $f$:
\begin{align}
    f(\boldsymbol{H}_{A_{t}}, \boldsymbol{U}_i) = \boldsymbol{h}^T \text{ReLU} (\boldsymbol{W_1}[\boldsymbol{H}_{A_{t}} \oplus \boldsymbol{U}_i] + \boldsymbol{b}).
\end{align}
In this \ac{MLP}, we first use the weight matrix $\boldsymbol{W_1}$ and bias vector $\boldsymbol{b}$ to map the input into a hidden layer, and then projects it into an attention weight via the weight vector $\boldsymbol{h}^T$.

Then, we can get the account representation by the following attention network:
\begin{align}
    & \boldsymbol{G}_c^A = \sum_{i=1}^{K_A} \alpha_{i,t} \boldsymbol{U}_i, \\
    & \alpha_{i,t} = \frac{\text{exp}(f(\boldsymbol{H}_{A_{t}}, \boldsymbol{U}_i))}{\sum_{i=1}^{K_A}(\text{exp}(f(\boldsymbol{H}_{A_{t}}, \boldsymbol{U}_i))^\beta}.
\end{align}
Inspired by~\cite{he2018nais}, we do not perform the standard softmax function on attention weights. Instead, we introduce the smoothing component $\beta\in [0.1,1]$ to avoid overly punishing the weights of accounts with many shared users. Obviously, when setting $\beta$ to 1, it can recover the softmax function; otherwise, it can suppress the value of denominator to avoid over punishing.

\subsubsection{Recommendation with Cross-domain Information}
\noindent Up to now, we only learn the account representation from one domain.
To make cross-domain recommendations, we need to transfer users' interactions from domain B to A (take A as the target domain), and then combine them with the original information in A to improve the recommendation in A. The transferred information from B to A is captured by another \ac{UIN}, which takes the whole transferred sequence as input, and outputs the corresponding account representation (denoted as $\boldsymbol{G}_c^{B\to A}$) with the target item in A as the supervision signal.
To integrate the information from both domains, we further combine (the concatenation operation is applied) $\boldsymbol{G}^A_c$ with the transferred knowledge ($\boldsymbol{G}_c^{B\to A}$) to get a comprehensive account representation.
Finally, the probability of recommending item $A_m$ is achieved by matching the account representation with the item embedding:
\begin{align}
    P(A_{m}|S^o_c) = \text{sigmoid} (\boldsymbol{H}_{A_{m}} \cdot [\boldsymbol{G}_c^{B \to A}; \boldsymbol{G}_c^A]^T + \boldsymbol{b}),
\end{align}
where $S^o_c$ is the mixed behavior sequence from account $G_c$ in both domains,
and $\boldsymbol{b}$ is the bias term. 
Similar to the prediction model in domain A, the prediction probability of item $B_n$ in domain B can be formulated as:
\begin{align}
    P(B_{n}|S^o_c) = \text{sigmoid} (\boldsymbol{H}_{B_{n}} \cdot [\boldsymbol{G}_c^{A \to B}; \boldsymbol{G}_c^B]^T + \boldsymbol{b}),
\end{align}
where $\boldsymbol{G}_c^{A \to B}$ denotes the transferred information from A to B, and $\boldsymbol{G}_c^B$ is the original information in B.

\subsubsection{Objective Function}
\noindent As each user interaction is a binary value 1 or 0, we deem the learning of a next item recommendation model as a binary classification task. That is, we treat the observed user-item interactions as positive instances and sample the negative instances from unobserved user-item pairs (we set the number of negative samples per positive as 4 in experiments).
Then, we formulate our objective function as a regularized log loss:
\begin{align}
    & L_A(\Theta)= -\frac{1}{T} (\sum_{s\in \mathcal{S}^+}\sum_{A_m\in s} \text{log} P (A_{m+1}|s) \notag \\
    & \qquad \qquad +\sum_{s\in \mathcal{S}^-}\sum_{A_m\in s} \text{log} (1- P (A_{m+1}|s))) + \lambda ||\Theta||^2, \\
    & L_B(\Theta)= -\frac{1}{T}(\sum_{s\in \mathcal{S}^+}\sum_{B_n\in s} \text{log} P (B_{n+1}|s) \notag \\
    & \qquad \qquad + \sum_{s\in \mathcal{S}^-}\sum_{B_n\in s} \text{log} (1 - P (B_{n+1}|s)))
    + \lambda ||\Theta||^2,
\end{align}
where $T$ denotes the total number of training instances, $\Theta$ represents all the trainable parameters, $\mathcal{S}^+$ and $\mathcal{S}^-$ denote the set of positive and negative instances, respectively.

% To enhance the process in cross-domain scenario, we jointly train $L_A$ and $L_B$ on both domains as follows:
% \begin{align}
%     L (\Theta) = L_A (\Theta) + L_B(\Theta).
% \end{align}

All the parameters in $L_A(\Theta)$ and $L_B(\Theta)$ are learned by a variant stochastic gradient descent method Adagrad~\cite{duchi_adaptive_2011} that applies an adaptive learning rate for each parameter and adopts the mini-batch to speedup the training process.
Note that, to avoid the seesaw phenomenon leading by a joint loss function that the approach often improves a domain at the sacrifice of another domain's performance, we exploit a separated objective function with a separated optimizer for each domain.

\subsection{Reinforcement Learning-enhanced domain Filter\label{domain_adapter}}
\noindent Though the cross-domain information can help us capture users' domain-specific preferences, the recommender might be disturbed if the transferred information is irrelevant or event opposite, which may happen when an account is used by users with different interests in two domains. For example, in a TV family account, children like the animation channel, while parents prefer the movie channel. Taking all the cartoons into account will not improve the performance of predicting the next movie for a parent. Instead, the irrelevant interactions dilute the main intent of the current behavior sequence. To deal with this issue, we develop a reinforcement learning-enhanced domain filter to refine the transferred interactions from the source domain to improve the recommendation performance in the target domain. Rather than assigning attention coefficients to each of the transferred behaviors, we propose to enhance the cross-domain recommender by removing the noisy ones. In the following, we will present its design and objective function.

\subsubsection{Learning Framework}
\noindent Due to the unavailability of direct supervision information, it is hard to leverage traditional methods to directly distinguish noisy interactions. Inspired by~\cite{Zhang2019HierarchicalRL, wang2018reinforcement}, we cast the task of domain filtering as a hierarchical \ac{MDP}, and decompose the overall task into two sub-tasks $T^h$ and $T^l$, where $T^h$ is a high-level task with one binary action determining 
whether to revise the whole transferred sequence, and $T^l$ is a low-level task with multiple actions deciding the necessity of removing each transferred interaction within it.
% We design a high-level task to determine whether to revise the whole transferred sequence or not, because a part of transferred sequences are already have efficient information to improve the basic cross-domain recommendation.
% to determine whether to remove each transferred interaction or not.
% Note that, the low-level task is executed if and only if the high-level task decides to make revisions, otherwise the overall task will be directly finished. 
Each kind of task is defined as a 4-tuple \ac{MDP} $(\mathcal{S}, \mathcal{A}, \mathcal{T}, \mathcal{R})$, where
\begin{itemize}
    \item $\mathcal{S}$: is a set of states defined in a continuous feature space;
    \item $\mathcal{A}$: is a set of actions that can be adopted by a whole sequence or every transferred interaction;
    \item $\mathcal{T}$: $\mathcal{S}\times \mathcal{A} \times \mathcal{S} \to [0,1]$ is the state transition probability mapping $\mathcal{S}\times \mathcal{A} \times \mathcal{S}$ into $[0,1]$;
    \item $\mathcal{R}$: $\mathcal{S}\times \mathcal{A} \to \mathbb{R}$ is the reward function mapping $\mathcal{S}\times \mathcal{A}$ into a real value.
\end{itemize}

Take the recommendation scenario in domain B as an example. Given an account $G_c$, an interaction sequence transferred from domain A to B ($\mathcal{S}_c^{A\to B}:=({A_1}^{A\to B}, {A_2}^{A\to B}, ..., {A_m}^{A\to B}, ...)$), and the target item $B_t$ in B, the agent (i.e., domain filter) performs a high-level task to decide whether to revise $\mathcal{S}_c^{A\to B}$ or not. If it does, the agent further conducts a low-level task to decide whether to remove each interaction $A_{m}^{A\to B}\in \mathcal{S}_c^{A\to B}$ or not.
Otherwise the overall task will be directly finished.
% filtering task to determine whether to remove each transferred interaction ($\boldsymbol{h}_{i}^{A\to B}\in \mathcal{H}_c^{A\to B}$) from the source domain. 
Then, the refined sequences are further fed into \ac{BCR}. To describe the framework of our agent (as shown in Fig.~\ref{fig:framework}) more clearly, we follow the common terminologies in reinforcement learning~\cite{williams1992simple}:

\textbf{Environment:} In our design, the environment consists of the account set $\mathcal{G}$, the item set $\mathcal{V}$, the basic cross-domain recommender, and the cross-domain interactions to be refined.

\textbf{States:} 
In the high-level task, the agent takes an action according to the state of the whole transferred sequence. Motivated by~\cite{Zhang2019HierarchicalRL}, we define the state $\boldsymbol{S}^h$ as: 1) the average cosine similarity between the embeddings of each transferred interaction $\boldsymbol{H}_{A_m}^{A\to B} \in \boldsymbol{H}_c^{A\to B}$ and the target item in domain B ($\boldsymbol{H}_{B_t}$), 2) the average element-wise product between the embeddings of $\boldsymbol{H}_{A_m}^{A\to B}$ and $\boldsymbol{H}_{B_t}$, and 3) the probability of recommending $B_t$ to account $G_c$ by the basic recommender based on $\boldsymbol{H}_c^{A \to B}$, where a lower recommendation probability indicates that we should take more efforts to revise $\mathcal{S}_c^{A\to B}$.

In the low-level task, the agent takes a sequence of actions according to the state $\boldsymbol{S}_m^l$ of each transferred interaction/item. Similar to $\boldsymbol{S}^h$, the state $\boldsymbol{S}_m^l$ is defined as: 1) the cosine similarity between the representation of the current item $\boldsymbol{H}_{A_m}^{A\to B}$ and the target $\boldsymbol{H}_{B_t}$, 2) the average value of 1) on all the reserved interactions, 3) the absolute value of the difference 
between $\boldsymbol{H}_{A_m}^{A\to B}$ and $\boldsymbol{H}_{B_t}$, and 4) the average value of 3).
% of the two previous features over all the reserved domain sequence.
The representation of an item or a transferred interaction is learned by \ac{BCR}.
% A state $s\in \mathcal{S}$ is the feature representations of account $g$'s historical interactions, transferred domain knowledge and the target item: $s=(h^B_g, h_g^{A\to B}, v_j)$.

\textbf{Action and policy:} We define the high-level action as a binary value $a^h\in \{0, 1\}$ to indicate whether the transferred sequence should be revised, and a low-level action as a binary indicator to decide whether the current transferred item should be removed, which are instructed by the following policy functions. 

The high-level policy function for a given sequence $\mathcal{S}_c^{A\to B}$ is defined as:
\begin{align}
    & \pi (\boldsymbol{S}^h, a^h) = P(a^h|\boldsymbol{S}^h, \Phi^h) \notag \\
    &\qquad = a^h \sigma(\boldsymbol{W}_1^h\boldsymbol{E}^h) + (1-a^h)(1-\sigma(\boldsymbol{W}_1^h\boldsymbol{E}^h)),
\label{Eq:high_policy}
\end{align}
where $\boldsymbol{W}_1^h\in \mathbb{R}^{d_1 \times d_2}$ is the parameter to be learned with dimension $d_1$ (the number of the state features) and $d_2$ (the dimension of the hidden layer). $\sigma (x)$ is the sigmoid function used to transform the input into a probability. $\boldsymbol{E}^h$ represents the embedding of the input state, which is defined as:
\begin{align}
    \boldsymbol{E}^h = \text{ReLU}(\boldsymbol{W}_2^h \boldsymbol{S}^h + \boldsymbol{b}^h),
\end{align}
where $\boldsymbol{W}_2^h \in \mathbb{R}^{d_2 \times 1}$ and $\boldsymbol{b}^h\in \mathbb{R}^{d_2}$ are the weight and bias of a neural network. We use $\Phi^h$ to denote the parameter set $\Phi^l = \{\boldsymbol{W}_1^h, \boldsymbol{W}_2^h, \boldsymbol{b}^h\}$ used in high-level policy function.

% The action $a_i\in \{0, 1\}$ representing whether to filter out the current transferred interaction $\boldsymbol{h}_i^{A\to B}$ from domain A to B is defined as a binary value.
The low-level policy function that instructs how we perform action $a_m^l$ is defined as:
\begin{align}
    & \pi (\boldsymbol{S}_m^l, a_m^l) = P(a_m^l|\boldsymbol{S}_m^l, \Phi^l) \notag \\
    &\quad \quad \quad= a_m^l \sigma(\boldsymbol{W}_1^l\boldsymbol{E}_m^l) + (1-a_m^l)(1-\sigma(\boldsymbol{W}_1^l\boldsymbol{E}_m^l)), \\
    & \quad \boldsymbol{E}_m^l = \text{ReLU}(\boldsymbol{W}_2^l \boldsymbol{S}_m^l + \boldsymbol{b}^l),
\label{Eq:low_policy}
\end{align}
where $\Phi^l = \{\boldsymbol{W}_1^l, \boldsymbol{W}_2^l, \boldsymbol{b}^l\}$ is the parameter set in the low-level policy function, $\boldsymbol{E}_m^l$ represents the embedding of the input state $\boldsymbol{S}_m^l$, $\boldsymbol{W}_1^l\in \mathbb{R}^{d_1 \times d_2}, \boldsymbol{W}_2^l \in \mathbb{R}^{d_2 \times 1}$ and $\boldsymbol{b}^l\in \mathbb{R}^{d_2}$ are the weights and bias of a neural network.

\textbf{Reward:} The reward is a signal indicating the reasonableness of the performed actions. For a low-level task, we assume every performed low-level action has two kinds of rewards, i.e., the immediate reward and the delayed long-term reward. We exploit a combined reward mainly because the immediate reward can speed up the agent's training process for the low-level action by giving every performed action an immediate feedback. Without the immediate reward, the low-level actions will only receive the same delayed long-term reward after the last low-level action is performed, which will 
% probably 
lead to slow local convergence. The detailed definitions of the rewards are shown as follows.
% The reward is a signal indicating the reasonableness of the performed actions,
% % \textcolor{red}{Inspired by~\cite{Zhang2019HierarchicalRL}, we design the rewards for high-level tasks and low-level tasks respectively. The details are as follows:}
% that is, whether it is beneficial to remove the transferred interactions.
% For state $s=(c^A, v_i, v_j)$ and action $a$, 
% Inspired by~\cite{Zhang2019HierarchicalRL}, 
% For a low-level task, we assume every performed low-level action has two kinds of rewards, i.e., the immediate reward and the delayed long-term reward.

% \textcolor{red}{\textbf{Reward for low-level actions:} The reward for the low-level action is a combination of the immediate reward and the long-term reward., where the long-term reward measures the long-term gain of transferring the cross-domain knowledge after the last action $a_{l_a-1}$ is performed for the last item of the sequence and the immediate reward measures the difference between after and before the transferred interaction is considered in the target domain, which is calculated by the average cosine similarity between the reserved items and the target item in B after and before the transferred item $A_m^{A\to B}$ is saved. }

% \textcolor{red}{Specifically, the  calculation of the immediate reward for each performed action is formulated as:}
\textbf{The immediate reward} measures the difference brought by considering the
% the difference between after and before the 
transferred interaction in the target domain, which is measured by the changes in the average
% calculated by the average
cosine similarity between the reserved items and the target item in B after
% and before
the transferred item $A_m^{A\to B}$ is saved. The calculation for each performed action is formulated as:
\begin{align}
    % R^I(a_t, s_t) = c_t\cdot v_i - c_t \cdot v_i,  
  & R^I(a_m^l, \boldsymbol{S}_m^l) = \frac{1}{|\mathcal{S}_c^{A\to B}|}\sum_{A_i\in \mathcal{S}_c^{A\to B}} \text{cosine}(\boldsymbol{H}_{A_i}^{A\to B}, \boldsymbol{H}_{B_t})  \notag \\
     & \quad \quad \quad \quad - \frac{1}{|\hat{\mathcal{S}}_c^{A\to B}|}\sum_{A_i\in \hat{\mathcal{S}}_c^{A\to B}} \text{cosine}(\boldsymbol{H}_{A_i}^{A\to B}, \boldsymbol{H}_{B_t}),
\end{align}
where $R^I(a_m^l,\boldsymbol{s}_m^l)$ is the immediate reward of performing $a_m^l$ given $\boldsymbol{s}_m^l$, $\boldsymbol{H}_{B_t}$ is the representation of the target item in domain B, $\boldsymbol{H}_c^{A\to B}$ is the embeddings of the transferred interactions, and $\hat{\mathcal{S}}_c^{A\to B}$ is the revised behavior sequence, which is a subset of $\mathcal{S}_c^{A\to B}$ excludes item $A_m^{A\to B}$. The immediate reward encourages the agent to keep the transferred items that are relevant to the target.
% As every performed action can get an immediate feedback, this design is proposed to speed up the agent's training process.

% \textcolor{red}{And the delayed long-term reward $R^D(a_m^l, \boldsymbol{S}_m^l)$ can be formulated as: }
\textbf{The delayed long-term reward} measures the long-term gain of transferring the cross-domain knowledge after the last action $a_{l_a-1}$ is performed for the last item. In other cases, its value is set to 0.
The formulation of the delayed long-term reward $R^D(a_m^l, \boldsymbol{S}_m^l)$ is shown as follows:
$$ R^D(a_m^l, \boldsymbol{S}_m^l)\!=\!\left\{
\begin{aligned}
  \text{log}p(\hat{\mathcal{S}}_, A_m) \!- \! \text{log} p(\mathcal{S}_c, A_m), \text{if } m= l_a-1;
  \\ \text{0\qquad\qquad,\qquad}\text{otherwise},
\end{aligned}
\right.
$$
where $p(\mathcal{S}_c, A_m)$ is an abbreviation of $p(y=1|\mathcal{S}_c^{A \to B}, A_m^{A\to B})$ denoting the positive instance of recommending item $A_m^{A\to B}$. We omit the superscript $A\to B$ due to space constraints. A positive reward of the above function indicates a positive utility gained by the revised profile. For the special case that all the transferred items are removed, we just set $R^D(a_m^l, \boldsymbol{S}_m^l)=0$, and only use the original information in domain B for recommendation.

To consider both the immediate and the delayed long-term reward, we sum $R^I$ and $R^D$ as the final reward function for the low-level action:
% \textcolor{red}{So that, the final reward function for the low-level task can be summed as:}
\begin{align}
    R(a_m^l, \boldsymbol{S}_m^l) = R^I(a_m^l, \boldsymbol{S}_m^l) + R^D(a_m^l, \boldsymbol{S}_m^l).
\label{eq:reward}
\end{align}
% \textcolor{red}{Note that, in our combined reward, the delayed long-term reward is a signal to indicate whether the performed actions are reasonable or not, and the immediate reward encourages the agent to keep the transferred items that are relevant to the target item.}

% \textcolor{red}{\textbf{The reward for high-level actions:} }
For the high-level action, it obtains the same delayed long-term reward as the low-level action, if the high-level task chooses to revise the transferred domain information. Otherwise, it keeps the original domain knowledge and receives a zero reward.

\begin{algorithm}[t]
%\small
\caption{\fontsize{9.4bp}{17bp}$\proc{The Overall Training Process of 
\ac{RL-ISN}}.$}
\label{alg:training}
\begin{algorithmic}[1]
\STATE Pre-train the basic cross-domain recommendation model;
\STATE Pre-train the domain filter by running Algorithm \ref{alg:domain_filter} with the basic recommender fixed;
\STATE Jointly train the basic recommender and domain filter by running Algorithm
\ref{alg:domain_filter};
\end{algorithmic}
\end{algorithm}

\begin{algorithm}[t]
%\small
\caption{\fontsize{9.4bp}{17bp}$\proc{The training process of \ac{RL-DF}}.$}
\label{alg:domain_filter}
\begin{algorithmic}[1]
\REQUIRE Training data $\mathcal{S}:=\{ \mathcal{S}_1^o, \mathcal{S}_2^o, ..., \mathcal{S}_c^o, ...\}$, a pre-trained basic cross-domain recommender and a domain filter parameterized by $\Theta$ and $\Phi$, respectively;
\ENSURE Model parameters $\Theta$ and $\Phi$;
\STATE Initialize $\Theta$ and $\Phi$;   
\FOR{episode $i=1$ to $I$}
  \FOR{$\mathcal{S}_c^o$ in $\mathcal{S}$}
  \item Sample a high-level action $a^h$ with $\Phi^h$;
  \IF {$a^h=0$}
    \item $R(a^h, \boldsymbol{S}^h)=0$;
  \ELSE
%   \FOR{$A_m$ (or $B_n$) in $\mathcal{S}_c^o$}
    \item Sample a sequence of low-level actions $\{a_1^l, a_2^l, ..., a_{t_s}^l\}$ with $\Phi^l$ for the cross-domain sequence;
    \item Compute $R(a_{m}^l, \boldsymbol{S}_{m}^l)$ by Eq. (\ref{eq:reward});
    \item Compute gradients by Eq. (\ref{eq:gradients_1}) and Eq. (\ref{eq:gradients_2});
%   \ENDFOR
  \ENDIF
   \ENDFOR
   \item Update $\Phi$ by gradients;
   \item Update $\Theta$ in the basic cross-domain recommender;
\ENDFOR
\end{algorithmic}
\end{algorithm}

\subsubsection{Objective Function}
\noindent To find out the optimal parameters of policy functions, we maximize the following expected reward:
\begin{align}
    \Phi^*=\text{argmax}_\Phi \sum_\tau P_\Phi(\tau; \Phi) R(\tau), 
\end{align}
where $\Phi =\{\Phi^l, \Phi^h\}$ represents the parameter set defined in policy functions, $\tau$ is a sequence of the sampled actions and the transited states, which can be denoted as $\{\boldsymbol{S}^h, a^h\}$ for the high-level task, and $\{\boldsymbol{S}_1^1, a_1^l, \boldsymbol{S}_2^l, a_2^l, ..., \boldsymbol{S}_m^l, a_m^l, ...,\boldsymbol{S}_{t_s}^l, a_{t_s}^l\}$ for the low-level task. $R(\tau)$ is the reward for the sampled sequence $\tau$, and $P_\Phi (\tau; \Phi)$ denotes the probability of sampling $\tau$ with $\Phi$. Since the sequences of these two tasks have too many possible action-state trajectories, we exploit the policy gradient theorem~\cite{sutton2000policy} and monto-carlo based policy gradient method~\cite{williams1992simple} to sample $J$ action-state trajectories, and calculate the gradients of the parameters based on that. 

The corresponding gradient of the parameters for the high-level task is calculated as:
\begin{align}
    \bigtriangledown_{\Phi^h}= \frac{1}{J} \sum_{j=1}^J\bigtriangledown_{\Phi^h}\text{log} \pi_{\Phi^h}(\boldsymbol{S}_{j}^h, a_{j}^h) R(a_j^h, \boldsymbol{S}_j^h),
\label{eq:gradients_1}
\end{align}
where the reward $R(a_j^h, \boldsymbol{S}_j^h)$ is assigned as the low-level reward of the last action $R(a_{t_s,j}^h, \boldsymbol{S}_{t_s,j}^h)$, and 0 otherwise.

The gradient of the parameters for the low-level task is presented as:
\begin{align}
    \bigtriangledown_{\Phi^l}\!=\! \frac{1}{J} \!\sum_{j=1}^J\!\sum_{m=1}^{t_s}\bigtriangledown_{\Phi^l}\text{log} \pi_{\Phi^l}(\boldsymbol{S}_{m,j}^l, a_{m,j}^l) R(a_{m,j}^l, \boldsymbol{S}_{m,j}^l),
\label{eq:gradients_2}
\end{align}
where $R(a_{m,j}^l, \boldsymbol{S}_{m,j}^l)$ is the reward of each action-state pair in sequence $\tau^{(j)}$.

\subsection{Model Training}
\noindent As the domain filter and the cross-domain recommender are interleaved together, we need to train them jointly. The training process is shown in Algorithms \ref{alg:training} and \ref{alg:domain_filter}, where we first pre-train the basic cross-domain recommender based on the original dataset. Then, we fix the parameters of \ac{BCR} and pre-train the domain filter to automatically revise the transferred interactions. Finally, we jointly train the two models together. To have a stable update, we follow the method in~\cite{Zhang2019HierarchicalRL,feng2018reinforcement}, and update each parameter by a linear combination of its old version and the new version, i.e., $\Phi_{new} = \lambda \Phi_{new } +(1-\lambda) \Phi_{old}$, where $\lambda \ll 1$.

\section{Experimental Setup}
% \noindent In this section, we first introduce the research questions to be answered in experiments, and then describe the datasets, evaluation methods, and baselines used in this work.

\subsection{Research Questions}
\noindent Our proposals are fully evaluated by answering the following research questions.
\begin{itemize}
    \item[\textbf{RQ1}] How does our proposed \ac{RL-ISN} perform compared with other state-of-the-art cross-domain recommenders?
    \item[\textbf{RQ2}] What are the performances of \ac{RL-ISN} on different domains? Is it helpful to leverage the cross-domain information?
    \item[\textbf{RQ3}] 
    % Is it helpful to model the shared-account characteristic? How does the User Identification Network (UIN) contribute to the performance of \ac{RLIS}?
    How do the key components of \ac{RL-ISN}, i.e., User Identification Network (UIN), \ac{RL}-based Domain Filter (DF), and positional encoding contribute to the recommendation performance?
    % Is it helpful to model the shared-account characteristic and refine the cross-domain information?
    \item[\textbf{RQ4}] 
    How do the hyper-parameters affect the performance of \ac{RL-ISN}?
    \item[\textbf{RQ5}] How is the training efficiency and scalability of \ac{RL-ISN} when processing large-scale data?
\end{itemize}

\begin{table}
    \centering
    \footnotesize
     \caption{Statistics of the datasets, where V, E, M and B represent the domains in the two datasets.}
    \begin{tabular}{l|c|c|c|c}
    \toprule
    \multicolumn{1}{c|}{\multirow{2}[1]{*}{\textbf{Details}}}&\multicolumn{2}{c|}{HVIDEO}&       \multicolumn{2}{c}{HAMAZON} \\
    \cmidrule{2-5}
    & V & E
    & M & B \\
    \midrule
    \#Items &16,431 &3,389 &67,161 & 126,547\\
    \#Interactions &1,970,378&2,052,966&4,406,924&4,287,240\\
    \midrule
    \#Accounts & \multicolumn{2}{c|}{13,714}
    & \multicolumn{2}{c}{13,724} \\
    \#Sequences & \multicolumn{2}{c|}{125,943}
    & \multicolumn{2}{c}{196,297} \\
    \#Training-targets & \multicolumn{2}{c|}{3,868}
    & \multicolumn{2}{c}{51,316} \\
    \#Test-targets & \multicolumn{2}{c|}{3,753}
    & \multicolumn{2}{c}{50,800} \\
    \bottomrule
    \end{tabular}
    \label{tab:dataset_statistics}
\end{table}

\subsection{Datasets}
\noindent We evaluate \ac{RL-ISN} on two real-world datasets HVIDEO~\cite{ma2019pi} and HAMAZON~\cite{ren2019net}.

HVIDEO is a smart TV dataset that records the watching logs of family accounts (shared-account) on two platforms, i.e., the V-domain and E-domain, from October 2016 to June 2017. The V-domain is a video watching platform containing TV series, movies, cartoons, etc. The E-domain includes the educational videos from elementary to high school, as well as instructional videos on sports, food, etc. As these records are generated by shared family accounts on two domains, they are suitable for investigating the \ac{SCSR} task. With further filtering out accounts that have less than 10 watching videos and those records with watching time less than 300 seconds, our resulting dataset contains 13,714 overlapped accounts and 125,943 sequences.

HAMAZON is a product review dataset released by Sun et al.~\cite{ren2019net}, which contains users' review behaviors on two Amazon platforms, i.e., M-domain and B-domain, from May 1996 to July 2014. The M-domain refers to Amazon users' watching and rating interactions on movies. The B-domain refers to users' reading and rating behaviors on books. This dataset satisfies the cross-domain characteristic via only keeping users who have interactions in both Amazon movie and book domains.
To simulates the shared-account characteristic, this dataset first splits the time schedule into 6 intervals, which are 1996-2000, 2001-2003, 2004-2006, 2007-2009, 2010-2012, 2013-2015. Then, it randomly merges 2-4 users and their review records in each interval as one shared-account. After splitting each sequence into small fragments within one year, and filtering out the sequences with less than 5 items from M-domain and 2 items from B-domain, the resulted dataset is finally reached, which contains 13,724 accounts and 196,297 sequences. The statistics of these two datasets are summarized in Table~\ref{tab:dataset_statistics}.

\begin{table*}
  \centering
  \scriptsize
   \caption{Comparison results (\%) on HVIDEO and HAMAZON. We do not report VUI-KNN on HAMAZON, as it needs specific time in a day which is not available in this dataset.}
    \begin{tabular}{lcccccccc|cccccccc}
    \toprule
    \multicolumn{1}{c}{\multirow{3}[4]{*}{\textbf{Methods}}} & \multicolumn{8}{c|}{\textbf{HVIDEO}} &
    \multicolumn{8}{c}{\textbf{HAMAZON }} \\
    % \midrule
    \cmidrule{2-17}
    & \multicolumn{4}{c}{\textbf{E-domain }} & \multicolumn{4}{c|}{\textbf{V-domain }} & \multicolumn{4}{c}{\textbf{M-domain}} & \multicolumn{4}{c}{\textbf{B-domain}} \\
    \cmidrule{2-17}
          & \multicolumn{2}{c}{HR} & \multicolumn{2}{c}{NDCG} 
          & \multicolumn{2}{c}{HR} & \multicolumn{2}{c|}{NDCG}
          &
          \multicolumn{2}{c}{HR} & \multicolumn{2}{c}{NDCG}
          & \multicolumn{2}{c}{HR} & \multicolumn{2}{c}{NDCG} 
          \\
          \midrule
          & @5      & @10   & @5    & @10   & @5    & @10   & @5    & @10  & @5     & @10   & @5    & @10   & @5    & @10   & @5    & @10
          \\
    \midrule
    % POP   &11.24&14.72&9.97&12.06&33.58&36.81&31.55&32.64
    % &5.84&6.92&5.11&5.78&7.62&7.89&7.33&7.40\\
    Item-KNN &10.99&15.72&9.97&13.41&33.77&36.21&32.88&35.19
    &5.55&6.32&5.08&5.80&6.84&7.52&6.30&6.98\\
    BPR-MF &7.62&10.02&7.30&9.41&29.58&31.11&29.55&30.31
    &5.88&6.94&5.72&6.57&5.43&5.88&5.12&5.24\\
    NCF &22.54&26.62&15.67&19.88&70.12&70.84&66.28&67.50 
    &25.57&31.03&19.88&24.92&11.39&14.62&9.48&13.22\\
    Light-GCN &66.16&77.79&51.42&55.20&92.33&93.41&90.68&91.03
    &41.04&53.38&29.79&33.78&56.31&66.57&19.09&28.88\\
    \midrule
    VUI-KNN &15.77&18.92&13.32&15.48&47.42&51.54&43.21&46.78  
    &  -     &   -    &   -    &  -     &  -     &  -     &   -    &  - \\
    % MISS &   &      &       &   &          &      &          &  &            &      &       &   &     &     &   & \\
    \midrule
    NCF-MLP++ &25.44&30.47&21.43&23.52&70.15&72.40&66.44&68.52
    &29.08&37.30&27.03&30.67&16.70&21.55&13.33&15.13\\
    Conet &32.83&48.99&26.51&32.83&70.23&73.28&70.09&73.27
    &34.43&49.15&32.44&43.72&18.12&27.20&12.14&15.06\\
    \midrule
    GRU4REC&71.92&83.28&57.16&60.85&92.88&94.06&92.18&92.56
    &39.99&53.11&28.80&33.03&81.87&85.20&79.79&80.86\\
    HGRU4REC &73.60&84.43&59.09&62.61&93.18&94.17&92.46&92.77
    &40.02&53.19&28.64&32.89&80.58&83.50&79.05&79.99\\
    NAIS &62.16&67.43&54.44&56.52&90.40&91.43&88.55&90.23
    &38.11&46.86&31.62&33.80&77.80&80.44&72.22&73.10 \\

    \midrule
    $\pi$-net &73.92&84.57&59.58&63.04&93.74&94.69&92.71&93.02
   				 & 43.85 &56.77&31.76&35.94&83.87&86.16 &81.21&81.95\\
    PSJ-net &74.35&84.96&59.97&63.43&93.82&94.72&92.76&93.05
    &42.66&54.70&31.21&35.07&82.45&85.67&81.32&82.11 \\
    DA-GCN &75.64&84.67&63.47&66.40&93.17&94.23&92.04&92.38
    &43.16&55.93&31.42&35.54&81.94&85.02&80.17&81.16\\
    \midrule
    
    ISN-RL &\textcolor{black}{78.82}&\textcolor{black}{89.03}&\textcolor{black}{62.17}&\textcolor{black}{66.32}&\textcolor{black}{94.35}&\textcolor{black}{94.53}&\textcolor{black}{88.44}&\textcolor{black}{89.01}
    &\textcolor{black}{55.07} &\textcolor{black}{70.76} &\textcolor{black}{41.23} &\textcolor{black}{46.29} &\textcolor{black}{84.06} &\textcolor{black}{86.35} &\textcolor{black}{81.21} &\textcolor{black}{82.76}\\

     \ac{RL-ISN} &\textbf{83.97}    & \textbf{90.91}  & \textbf{79.22}    & \textbf{79.86}  & \textbf{94.55}     & \textbf{97.30}  & \textbf{93.12}      & \textbf{94.15}   
     &\textbf{56.71}&\textbf{73.28} &\textbf{42.08} &\textbf{47.44} &\textbf{89.48} &\textbf{94.61} &\textbf{83.31} &\textbf{84.97}   \\

    \bottomrule
    \end{tabular}%
  \label{tab:resutls}%
\end{table*}%

We take HAMAZON as a supplement to the real-life dataset mainly because: 1) We cannot find other datasets that contain shared-account information.
Though the CAMRa 2011 dataset~\cite{verstrepen2015top} also contains real-life shared-account information, the owner does not wish to distribute the dataset anymore.
2) The HAMAZON dataset has been commonly used for \ac{SCSR} in recent publications~\cite{guo_DAGCN_2021,ren2019net} as a supplement to the real-life dataset. In this work, we follow their experimental settings, and utilize the synthetic HAMAZON data to evaluate our method. Moreover, we also notice that using synthetic accounts is a common practice in studying the shared-account recommendation problem~\cite{verstrepen2015top, zhang2012guess, jiang2018identifying}. For example, Verstrepen et al.~\cite{verstrepen2015top} study the top-n recommendation for shared-account with synthetic accounts that are manually created by randomly merging several users' histories together. Jiang et al.~\cite{jiang2018identifying} follow the account creation strategy in~\cite{verstrepen2015top}, and study a user identification task based on that.

\subsection{Evaluation Protocols}
\noindent We hold out the last two items in each sequence as the training and test target~\cite{Zhang2019HierarchicalRL}, respectively. And, we formulate each data sample as a sequence of historical interacted items paired with a target item. For example, for the training data, we leverage the last second item as the target, and the rest (excluding the last item) as historical items. To accelerate our training process, we further construct $n$ negative samples (we will fine tune this hyperparameter later) for each positive instance by replacing the target item with each of $n$ randomly sampled items~\cite{Zhang2019HierarchicalRL}. For the test data, we treat the last item in each sequence as the test target, and the corresponding items of the same data sample in the training set as the historical items. Our task is to predict the exact target from the possible candidate items.

We take the widely used Hit Ratio (HR@$N$) and Normalized Discounted Cumulative Gain (NDCG@$N$)~\cite{he2015trirank} as our evaluation metrics to evaluate the ability of our method in recommending Top-$N$ items for each test instance. Specifically, if a ground-truth item appears in the recommended list, we have a hit; otherwise, we have a miss. Then, we define HR as the ratio of hits over ground-truth items in the Top-$N$ list.
However, as HR is not position aware, it can not reflect the performance of our method in getting top ranks correct. Hence, we adopt NDCG to address this, which assigns higher scores to correct recommendations at top ranks. For both metrics, a higher value denotes a higher recommendation performance. In experiments, we evaluate each test instance with both metrics, and report their average values.

\subsection{Baseline Methods\label{baselines}}
\noindent We compare \ac{RL-ISN} with the following baseline methods from five categories: traditional, shared-account, cross-domain, sequential, and shared-account cross-domain sequential recommendations.

1) Traditional recommendations:
\begin{itemize}
    % \item POP~\cite{ma2019pi}: This is a very simple yet frequently used strong baseline, which always recommends the most popular items in the training set.
    \item Item-KNN~\cite{hidasi2016srnn}: This method recommends the items that are similar to the actual item, and defines the item-to-item similarity as the cosine similarity between the vector of their sequences. 
    % This is one of the most common item-to-item solutions that assumes others who viewed this item also viewed these ones.
    \item BPR-MF~\cite{hidasi2016srnn}: This is a traditional matrix factorization-based method. We apply it for \ac{SR} via representing the sequence feature vector by averaging the feature vectors of items that had occurred in the sequence so far.
    % 我尝试添加了说明
    \item NCF~\cite{hexiangnan2017ncf}: This method treats users as sequences, and learns their embeddings by a \ac{MLP}-based collaborative filtering mechanism.
    \item Light-GCN~\cite{He2020lightGCN}: This is a GCN-based collaborative filtering method, which learns sequence and item ebmeddings through aggregating the representations of their neighbors.
\end{itemize}

\begin{table*}
    \caption{\fontsize{10.5bp}{17bp}Ablation studies (\%) on HVIDEO and HAMAZON.}
    \centering
    \scriptsize % footnotesize就会比较大
    \begin{tabular}{lcccccccc|cccccccc}
    \toprule
    \multicolumn{1}{c}{\multirow{3}[4]{*}{\textbf{Methods}}} & \multicolumn{8}{c|}{\textbf{HVIDEO}} &
    \multicolumn{8}{c}{\textbf{HAMAZON }} \\
    % \midrule
    \cmidrule{2-17}
    & \multicolumn{4}{c}{\textbf{E-domain }} & \multicolumn{4}{c|}{\textbf{V-domain }} & \multicolumn{4}{c}{\textbf{M-domain}} & \multicolumn{4}{c}{\textbf{B-domain}} \\
    \cmidrule{2-17}
          & \multicolumn{2}{c}{HR} & \multicolumn{2}{c}{NDCG} 
          & \multicolumn{2}{c}{HR} & \multicolumn{2}{c|}{NDCG}
          &
          \multicolumn{2}{c}{HR} & \multicolumn{2}{c}{NDCG}
          & \multicolumn{2}{c}{HR} & \multicolumn{2}{c}{NDCG} 
          \\
          \midrule
          & @5      & @10   & @5    & @10   & @5    & @10   & @5    & @10  & @5     & @10   & @5    & @10   & @5    & @10   & @5    & @10
          \\
    \midrule
    ISN-P &\textcolor{black}{83.87}    & \textcolor{black}{90.84}  & \textcolor{black}{79.18}    & \textbf{79.94}  & \textcolor{black}{94.43}     & \textcolor{black}{97.08}  & \textbf{93.80}      & \textcolor{black}{94.11}   
     &\textcolor{black}{55.27}&\textcolor{black}{69.13} &\textbf{43.24} &\textcolor{black}{46.25} &\textcolor{black}{88.94} &\textcolor{black}{94.51} &\textcolor{black}{82.84} &\textcolor{black}{84.94}   \\
    % ISN-RL&77.46&88.02&62.35&65.79&94.17&94.75&88.23&88.42
    % &54.88&63.43&41.42&45.90&83.84&86.27&82.89&85.58\\ 2022-05-02 revised
    ISN-RL & 78.82 & 89.03 & 62.17 & 66.32 & 94.35 & 94.53 & 88.44 & 89.01
    & 55.07 & 70.76 & 41.23 & 46.29 & 84.06 & 86.35 & 81.21 & 82.76\\
    % ISN-UI  &65.77 &77.16 &51.39 &55.09 
    % &92.26 &93.19 &84.58 &84.87 
    % &41.55 &52.28 &29.87 &33.52  
    % &76.03 &82.84 &74.41 &75.02 \\ 2022-05-02 revised
    ISN-UI & 66.22 & 79.45 &57.32 & 63.28 & 91.45 & 92.07 & 83.86 & 86.90
    & 44.13 & 55.95 & 33.22& 34.91 & 79.80 & 82.33 & 76.72 & 75.97 \\
    % ISN &62.01&75.57 &46.05 &50.45  
    % &92.20 &93.23 &91.88 &92.21
    % &41.31 &51.60 &30.12 &31.18  
    % &72.20 &80.14 &70.11 &73.52 \\ 2022-05-02 revised
    ISN & 62.45 & 75.03 & 49.80 & 55.13 & 91.42 & 92.65 & 85.92 & 86.94
    & 41.04 & 49.53 & 28.22 & 30.12 & 72.44 & 79.51 &71.18 & 72.02 \\
    \midrule
    \ac{RL-ISN} &\textbf{83.97}    & \textbf{90.91}  & \textbf{79.22}    & 79.86  & \textbf{94.55}     & \textbf{97.30}  & 93.12      & \textbf{94.15}   
     &\textbf{56.71}    & \textbf{73.28}  & 42.08   & \textbf{47.44}  & \textbf{89.48}     & \textbf{94.61}  & \textbf{83.31}      & \textbf{84.97}   \\
    \bottomrule
    \end{tabular}%
  \label{tab:ablations}%
\end{table*}%

2) Shared-account recommendations:
\begin{itemize}
    \item VUI-KNN~\cite{wang2014user}: This method is proposed for IP-TV recommendation task that first cuts the logs of each account into slices via dividing a day into three time periods; each log slice is assumed to be generated by a virtual user. Then, latent users are further formed by merging virtual users according to their cosine similarities. After that, the User-KNN method is applied to make recommendations for the latent users.
    % \item MISS~\cite{wen_miss_2021}: This method proposes a multi-user identification module to first distinguish different user behaviors under the same account via an attention mechanism, and then learn the account's representation according to the attention weight of these behaviors. 
\end{itemize}

Note that, we do not compare with the shared-account recommendation methods~\cite{li2009can,verstrepen2015top,yang2017personalized} that need extra information, such as explicit ratings or textual descriptions for items, which are not available in our datasets.

3) Cross-domain recommendations:
\begin{itemize}
    % 我尝试在这里进行了修改
    \item NCF-MLP++~\cite{hexiangnan2017ncf}: This method is based on the traditional NCF, which makes recommendations on a single domain. We use it for \ac{CR} by sharing the collaborative filtering in different domains.
    \item Conet~\cite{hu2018conet}: This is a neural transfer model for \ac{CR}, which proposes a neural collaborative filtering network for information sharing.
\end{itemize}

4) Sequential recommendations:
\begin{itemize}
    \item GRU4REC~\cite{hidasi2016srnn}: This is one of the most representative sequential recommenders that exploits GRU to encode sequence and optimizes a ranking-based loss function.
    \item HGRU4REC~\cite{quadrana2017hrnn}: This method further improves GRU4REC via taking user's identity and auxiliary features into sequential recommendation. 
    \item NAIS~\cite{he2018nais}: This is an item-to-item collaborative filtering algorithm that uses an attention mechanism to distinguish the importance of different historical courses. But it is proposed for traditional \ac{SR}, and has limited ability to deal with the cross-domain and shared-account challenges.
    % \item NARM
\end{itemize}

5) Shared-account cross-domain sequential recommendations:
\begin{itemize}
    \item $\pi$-net~\cite{ma2019pi}: This is a state-of-the-art recommendation method for \ac{SCSR}, which devises a gating mechanism to transfer the information to other domains and addresses the shared-account challenge via learning user-specific representations.
    \item PSJ-net~\cite{ren2019net}: This method further improves $\pi$-net by changing the splitting and joining methods of learning the cross-domain representations.
    % is a newly produced recommendation method to address the challenge raised by \ac{SCSR}. This work learns role-specific representations and uses a gating mechanism to filter out the information of user roles that might be useful for another domain.
    \item DA-GCN~\cite{guo_DAGCN_2021}: This is another recent state-of-the-art recommendation method for \ac{SCSR}, which leverages a domain-aware graph convolutional network to model the cross-domain knowledge. For the shared-account challenge, this work also learns user-specific node representations as its solution.
    \item ISN-RL: This is a variant of \ac{RL-ISN} that removes the fine tuning process within it, that is, removing the \ac{RL}-based domain filter component. This is to demonstrate the effectiveness of our \ac{RL}-based knowledge transfer process. 
\end{itemize}

\subsection{Implementation Details}
\noindent We implement \ac{RL-ISN} using Tensorflow accelerated by NVIDIA RTX 2080 Ti GPU. We initialize model parameters via the Xavier method~\cite{glorot2010understanding} and exploit Adam~\cite{kingma2014adam} to optimize our loss function. For the hyper-parameters in basic recommender, we set the item embedding as 16, the batch-size as 256, the number of negative samples per positive as 4, the learning rate as 0.01, the dropout rate as 0.1 at pre-training stage on both domains. In the joint-training stage, we set the learning rate as 0.0001, the delayed coefficient $\lambda$ as 0.0005. For the domain filtering agent, we set the sampling time $M$ as 3, the learning rate as 0.05 and 0.0001 at the pre-training state and joint stage, respectively. We set the delayed coefficient $\lambda$ as 0.0005. For the policy function, we set the dimensions of both hidden layers ($d_1$ and $d_2$) as 8. We search the number of users ($K_A$ and $K_B$) within $\{1, 2, 3, 4, 5\}$ per shared-account for both datasets.
For the hyper-parameter settings in comparative methods, we set their number of negative samples to 4 for the sake of fairness. For other parameters, we refer to the settings in their papers and also fine tune them on different datasets. 
We conduct the one sample paired t-tests to verify that all improvements are statistically significant for $p<0.01$.

% \begin{table*}[t]
%   \centering
%   \small
%   \caption{Case studies of the revised items by \ac{RL-ISN} and the cosine similarity scores on HVIDEO.}
%     \begin{tabular}{c|c|c}
%     \toprule
%     \multicolumn{1}{c|}{\textbf{Methods}} &
%     \multicolumn{1}{c|}{\textbf{Revised items or the attention scores }}  & \multicolumn{1}{c}{\textbf{The target item}}\\
%     % \midrule
%     \midrule
%     \ac{RL-ISN}&V289, V74, V10398, V326, V7638, V3032, V326
%     &E392\\
%     Basic Recommender&V289(19.82), V74(22.54), V10398(17.41), V326(23.44), V7638(22.71), V3032(19.55), V326(28.41)
%     &E392\\
%     \midrule
%     \ac{RL-ISN}&E241, E232, E80, E81, E28, E3
%     &V326\\
%     Basic Recommender &E241(22.63), E232(25.16), E80(16.73), E81(16.55), E28(21.45), E3(19.75)
%     &V326\\
%     \midrule
%     \ac{RL-ISN}&V4581, V926, V508, V137, V20
%     &E167\\
%     Basic Recommender &V4581(16.74), V926(28.14), V508(25.19), V137(23.08), V20(21.87)
%     &E167\\
%     \midrule
%     \ac{RL-ISN}&E39, E259, E23, E94, E163, E90
%     &V2961\\
%     Basic Recommender&E39(24.51), E259(19.22), E23(16.78), E94(18.22), E163(21.32), E90(17.91)
%     &V2961\\
%     \bottomrule
%     \end{tabular}%
%   \label{tab:case_studies}%
% \end{table*}%
\begin{figure*}
    \centering
    \includegraphics[width=18cm]{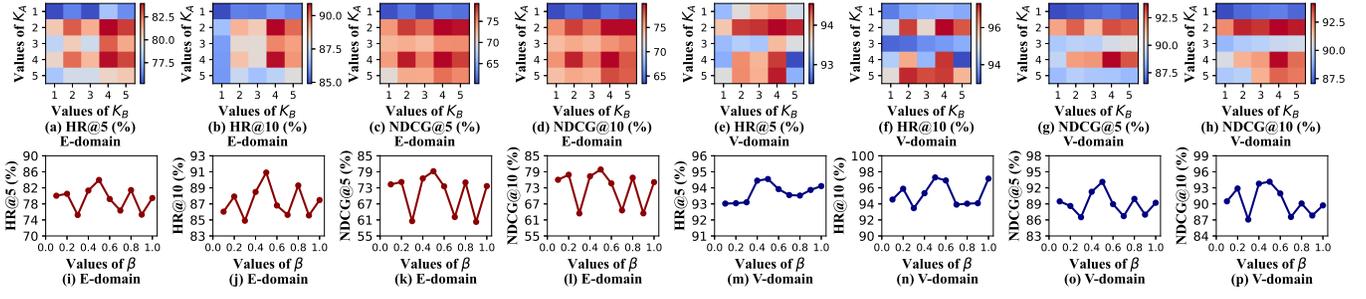}
    \caption{Impact of the hyper-parameters $K_A$, $K_B$ and $\beta$ on HVIDEO.}
    \label{fig:hyper_params_4_KB}
\end{figure*}
\section{Experimental Results (RQ1 \& RQ2)}
\noindent The comparison results on HVIDEO and HAMAZON are reported in Table~\ref{tab:resutls}, from which we have the following observations:
1) Our \ac{RL-ISN} method outperforms all the baselines on both datasets, demonstrating the advantage of \ac{RL-ISN} in modeling the shared-account representation and the cross-domain information, which leads to a better recommendation performance. Moreover, \ac{RL-ISN} achieves the best performance on both domains, demonstrating the capability of our solution in transferring the domain knowledge.
2) \ac{RL-ISN} outperforms the sequential recommendation methods that do not consider the cross-domain information (i.e, GRU4REC, HGRU4REC and NAIS) on both datasets, which again demonstrates the importance of the domain knowledge. The gap between \ac{RL-ISN} and traditional cross-domain recommendation methods (i.e., Conet and NCF-MLP++), indicating the effectiveness of solution in transferring the domain information.
3) \ac{RL-ISN} performs better than the methods that do not consider the shared-account characteristic (e.g., NAIS, Conet, and BPR-MF), showing the benefit of modeling an account from a shared-account view. Treating an account as a virtual user can not get better results than identifying them in advance.
4) The methods developed for \ac{SCSR} (i.e., $\pi$-net, PSJ-net, DA-GCN, ISN-RL and \ac{RL-ISN}) significantly outperform other baselines, denoting the benefit of simultaneously modeling the shared-account and cross-domain characteristics. Furthermore, the improvement of \ac{RL-ISN} over ISN-RL shows the necessity of fine tuning the cross-domain information, and the effectiveness of our \ac{RL}-based method.

\section{Model Analysis\label{analysis}}

\subsection{Ablation Studies (RQ3)}
\noindent To explore the importance of different components in \ac{RL-ISN}, we compare it with its following variants:
\begin{itemize}
    \item ISN-P (no Position): This is a variant of \ac{RL-ISN} that removes the positional encoding in the input layer. 
    This is to demonstrate the necessity of considering the position information in sequence modeling.
    \item ISN-RL (no RL): As shown in Section \ref{baselines}, this is a variant of \ac{RL-ISN} that removes the \ac{RL-DF} module within it.
    This is to demonstrate the effectiveness of our \ac{RL}-based fine tuning process. Note that, this method actually detergents to our basic recommender (i.e., \ac{BCR}).
    \item ISN-UI (no UI): This variant removes the user identification network form \ac{RL-ISN}. This is to evaluate the effectiveness of our attention-based user clustering process in modeling the shared-account characteristic.
    \item ISN (no RL, no UI, and no Position): This variant simultaneously removes the domain filter, user identification, and positional encoding components. This is to evaluate the performance improvement caused by incorporating them in \ac{RL-ISN}.
    
\end{itemize}

From the comparison results reported in Table~\ref{tab:ablations}, we have the following observations: 1) \ac{RL-ISN} outperforms ISN-RL on both datasets, demonstrating the importance of the \ac{RL-DF} component and the effectiveness of our RL-based fine tuning method. That is, treating the domain filtering as a hierarchical \ac{MDP} can help us find a more relevant domain information to the target domain. 2) \ac{RL-ISN} achieves a better result than ISN-UI, showing the benefit of modeling the shared-account characteristic and the utility of our user identification module. This result demonstrates that our model can learn a more expressive account representation via the attention mechanism over the latent users. 3) \ac{RL-ISN} outperforms ISN-P in terms of almost all metrics on both datasets, demonstrating the benefit of considering the position information in sequence modeling, and the effectiveness of our modeling method. 4) \ac{RL-ISN} performs better than ISN, indicating the importance of simultaneously consider the shared-account and cross-domain characteristics. And the attention-based user identification method as well as the \ac{RL}-based domain transferring process let us develop a more effectiveness recommendation model for \ac{SCSR}.

\subsection{Impact of Hyper-parameters (RQ4)}

\noindent This section investigates the impact of the parameters that are important to \ac{RL-ISN}. As we have similar results on HVIDEO and  HAMAZON, we only report the experimental results on HVIDEO as an example.

\textbf{Impact of Hyper-parameters $K_A$ and $K_B$.} In \ac{RL-ISN}, we devise a user identification module to model the shared-account characteristic via attentively aggregating the latent users sharing an account, and the numbers of the latent users are introduced as two hyper-parameters $K_A$ and $K_B$. To investigate how $K_A$ and $K_B$ affect the recommendation performance, we search them within $\{1, 2, 3, 4, 5\}$, and report the experimental results in Fig.~\ref{fig:hyper_params_4_KB}.
From the results, we can observe that the best performance of \ac{RL-ISN} is achieved when $K_A$ = 2 for domain A, and $K_B$ = 4 for domain B, which are consistent with the typical family sizes.
In other cases, \ac{RL-ISN} cannot reach a better performance, which further supports our motivation in modeling the shared-account characteristic.

\textbf{Impact of Hyper-parameter $\beta$.} To avoid too much punishment on the weights of accounts with many shared users, we introduced a smoothing hyper-parameter $\beta$. Fig.~\ref{fig:hyper_params_4_KB} shows the performance of  \ac{RL-ISN} with different $\beta\in[0.1,1]$. When setting $\beta$ to 1, it means we use a standard softmax to normalize the weights of \ac{RL-ISN}. From Fig.~\ref{fig:hyper_params_4_KB}, we find that when $\beta$ = 0.5, \ac{RL-ISN} can achieve its best performance, and it reaches worse performance when $\beta$ has bigger values,
proving the importance of our punishment strategy.

\begin{figure}
    \small
    \centering
    \includegraphics[width=8cm]{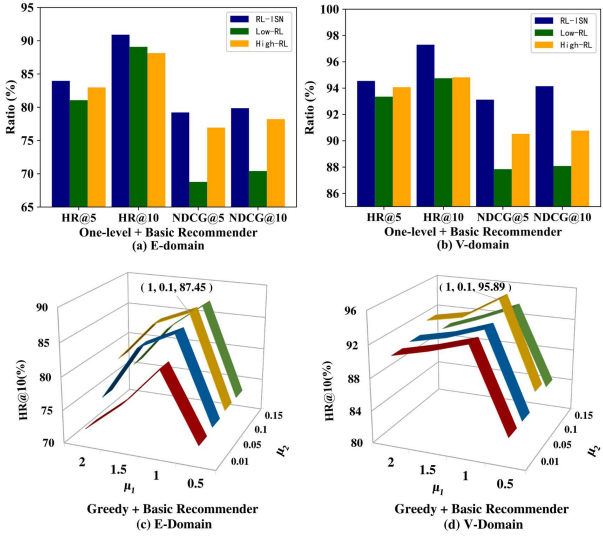}
    \caption{Model performance with one-level RLs and greedy revision algorithm on HVIDEO.}
    \label{fig:greedy}
\end{figure}

To further explore the utility of the \ac{RL}-based domain filter, we compare \ac{RL-ISN} with the following variants:
\begin{itemize}
    \item \textbf{Compared with One-level RLs}: We compare \ac{RL-ISN} with two variants of it, i.e., Low-RL and High-RL, where Low-RL denotes the method that only exploits the low-level RL to remove the noisy items from the transferred sequences, and High-RL is the method that only uses the high-level RL to decide whether to keep the transferred sequence or not. The comparison results are shown in Fig.~\ref{fig:greedy}, from which we observe that \ac{RL-ISN} outperforms Low-RL and High-RL on both domains, indicating the effectiveness of our hierarchical RL strategy, and having only leverage one-level RL is insufficient to get satisfactory results.
    \item \textbf{Compared with Greedy Revision}: We further compare \ac{RL-ISN} with a greedy revision method, which decides to revise whole sequence if $\text{log}p(\boldsymbol{S}_c^{A\to B})<\mu_1$ (suppose B is the target domain), and decides to remove an interaction $A_m^{A\to B}$ within the sequence if its cosine similarity is less than $\mu_2$. The experimental results are shown in Fig.~\ref{fig:greedy}, from which we can observe that the best performance (HR@10 = 87.45\%) is achieved on E-domain when $\mu_1 =1$ and $\mu_2=0.1$, which is 3.46\% less than \ac{RL-ISN} (1.41\% less than \ac{RL-ISN} on V-domain). This result demonstrates the importance of an appropriate revising strategy and the effectiveness of our RL-based domain filter.
\end{itemize}

% \begin{figure}[ht]
% \small
%     \subfigure{
%     \begin{minipage}{4.25cm}
%         \centering       
%         \includegraphics[width=4.25cm]{one_A.eps}
%     \end{minipage}
%     }
%     \subfigure{
%     \begin{minipage}{4.25cm} 
%         \centering
%         \includegraphics[width=4.25cm]{one_B.eps}
%     \end{minipage}
%     }
%     \subfigure{
%     \begin{minipage}{4.25cm}
%         \centering
%         \includegraphics[width=4.25cm]{greedy_c.eps}
%     \end{minipage}
%     }
%     \subfigure{
%     \begin{minipage}{4.25cm}
%         \centering
%         \includegraphics[width=4.25cm]{greedy_d.eps}
%     \end{minipage}
%     }
%     \caption{Model performance with one-level RL and greedy revision algorithm on HVIDEO.}
%     \label{fig:greedy}
% \end{figure}

\subsection{Training Efficiency and Scalability (RQ5)}

\begin{figure}[ht]
    \centering
    \small
    \includegraphics[width=8cm]{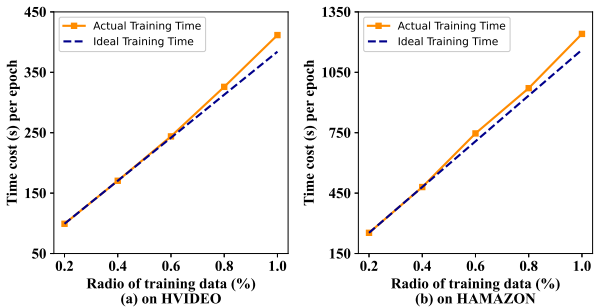}
    \caption{Training efficiency and scalability of \ac{RL-ISN}.}
    \label{fig:time_cost}
\end{figure}

\noindent To investigate the training efficiency and scalability of \ac{RL-ISN}, we validate them by measuring the time consumption of the training process with different data proportions on HVIDEO and HAMAZON. 
Fig.~\ref{fig:time_cost} shows the training cost with splitting the training data into $\{0.2, 0.4, 0.6, 0.8, 1.0\}$ while keeping all the hyper-parameters fixed.
To make our results comparable, we also report the expected ideal training time, i.e., the training cost is linearly associated with the training ratios, in Fig.~\ref{fig:time_cost}, from which we can find that the training cost almost linearly grows (from $0.102\times 10^3$ to $0.396\times 10^3$ on HVIDEO and from $0.294\times 10^3$ to $1.126\times 10^3$ on HAMAZON) with the increase of the training ratio on both datasets. This result provides us a positive answer to RQ5, that is, \ac{RL-ISN} is scalable to large-scale datasets.

\section{Conclusions and Future Work}
\noindent In this work, we investigate the \ac{SCSR} task, and propose a novel reinforcement learning-based solution, namely \ac{RL-ISN}. To simultaneously consider the shared-account and cross-domain characteristics, \ac{RL-ISN} develops an attention-based user identification network and a reinforcement learning-based domain filter, respectively. Then, to show the effectiveness of \ac{RL-ISN}, we conduct extensive experiments on two real-world datasets (i.e., HVIDEO and HAMAZON), and the experimental results demonstrate that our \ac{RL-ISN} solution is capable to well address the issues in \ac{SCSR} and can achieve better results than other state-of-the-art methods.

A limitation of \ac{RL-ISN} is that it assumes all accounts are shared by the same number of latent users, because in real-world scenarios, the number and identity of users under a shared-account are both unknown. That is, we can only know the origin of a behavior at the account-level, but not at the user-level.
We let the study of improving \ac{RL-ISN} via automatically detecting the number of latent users under a shared-account as our future work.
\ifCLASSOPTIONcompsoc
  % The Computer Society usually uses the plural form
  \section*{Acknowledgments}
\else
  % regular IEEE prefers the singular form
  \section*{Acknowledgment}
\fi
\noindent This work was supported by the National Natural Science Foundation of China (Nos. 61602282, 62072279), and Australian Research Council
(Nos. DP190101985, FT210100624).

% Can use something like this to put references on a page
% by themselves when using endfloat and the captionsoff option.
\ifCLASSOPTIONcaptionsoff
  \newpage
\fi

% trigger a \newpage just before the given reference
% number - used to balance the columns on the last page
% adjust value as needed - may need to be readjusted if
% the document is modified later
%\IEEEtriggeratref{8}
% The "triggered" command can be changed if desired:
%\IEEEtriggercmd{\enlargethispage{-5in}}

% references section

% can use a bibliography generated by BibTeX as a .bbl file
% BibTeX documentation can be easily obtained at:
% http://mirror.ctan.org/biblio/bibtex/contrib/doc/
% The IEEEtran BibTeX style support page is at:
% http://www.michaelshell.org/tex/ieeetran/bibtex/
%\bibliographystyle{IEEEtran}
% argument is your BibTeX string definitions and bibliography database(s)
%\bibliography{IEEEabrv,../bib/paper}
%
% <OR> manually copy in the resultant .bbl file
% set second argument of \begin to the number of references
% (used to reserve space for the reference number labels box)
% \begin{thebibliography}{1}

% \bibitem{IEEEhowto:kdfss}
% H.~Kopka and P.~W. Daly, \emph{A Guide to \LaTeX}, 3rd~ed.\hskip 1em plus
%   0.5em minus 0.4em\relax Harlow, England: Addison-Wesley, 1999.

% \end{thebibliography}

\bibliographystyle{IEEEtran}
\bibliography{sample-base}

% Generated by IEEEtran.bst, version: 1.14 (2015/08/26)
\begin{thebibliography}{10}
\providecommand{\url}[1]{#1}
\csname url@samestyle\endcsname
\providecommand{\newblock}{\relax}
\providecommand{\bibinfo}[2]{#2}
\providecommand{\BIBentrySTDinterwordspacing}{\spaceskip=0pt\relax}
\providecommand{\BIBentryALTinterwordstretchfactor}{4}
\providecommand{\BIBentryALTinterwordspacing}{\spaceskip=\fontdimen2\font plus
\BIBentryALTinterwordstretchfactor\fontdimen3\font minus
  \fontdimen4\font\relax}
\providecommand{\BIBforeignlanguage}[2]{{%
\expandafter\ifx\csname l@#1\endcsname\relax
\typeout{** WARNING: IEEEtran.bst: No hyphenation pattern has been}%
\typeout{** loaded for the language `#1'. Using the pattern for}%
\typeout{** the default language instead.}%
\else
\language=\csname l@#1\endcsname
\fi
#2}}
\providecommand{\BIBdecl}{\relax}
\BIBdecl

\bibitem{zhuang_cross_2010}
F.~Zhuang, P.~Luo, H.~Xiong, Y.~Xiong, Q.~He, and Z.~Shi, ``Cross-domain
  learning from multiple sources: {A} consensus regularization perspective,''
  \emph{TKDE}, pp. 1664--1678, 2010.

\bibitem{cao_transfer_2010}
B.~Cao, N.~N. Liu, and Q.~Yang, ``Transfer learning for collective link
  prediction in multiple heterogenous domains,'' in \emph{ICML}, 2010, pp.
  159--166.

\bibitem{abel2013cross}
F.~Abel, E.~Herder, G.~Houben, N.~Henze, and D.~Krause, ``Cross-system user
  modeling and personalization on the social web,'' \emph{UMUAI}, pp. 169--209,
  2013.

\bibitem{Zhu2021crossdomain}
F.~Zhu, Y.~Wang, C.~Chen, J.~Zhou, L.~Li, and G.~Liu, ``Cross-domain
  recommendation: Challenges, progress, and prospects,'' in \emph{AAAI}, 2021,
  pp. 4721--4728.

\bibitem{hu2018conet}
G.~Hu, Y.~Zhang, and Q.~Yang, ``Conet: Collaborative cross networks for
  cross-domain recommendation,'' in \emph{CIKM}, 2018, pp. 667--676.

\bibitem{Fan2021crossdomain}
W.~Fan, T.~Derr, X.~Zhao, Y.~Ma, H.~Liu, J.~Wang, J.~Tang, and Q.~Li,
  ``Attacking black-box recommendations via copying cross-domain user
  profiles,'' in \emph{ICDE}, 2021, pp. 1583--1594.

\bibitem{Salah2021crossdomain}
A.~Salah, T.~B. Tran, and H.~Lauw, ``Towards source-aligned variational models
  for cross-domain recommendation,'' in \emph{RecSys}, 2021, p. 176–186.

\bibitem{zhao2016passenger}
Y.~Zhao, J.~Cao, and Y.~Tan, ``Passenger prediction in shared accounts for
  flight service recommendation,'' in \emph{the 10th Conf. Asia-Pacific Serv.
  Comput.}, 2016, pp. 159--172.

\bibitem{jiang2018identifying}
J.-Y. Jiang, C.-T. Li, Y.~Chen, and W.~Wang, ``Identifying users behind shared
  accounts in online streaming services,'' in \emph{SIGIR}, 2018, pp. 65--74.

\bibitem{wen_miss_2021}
X.~Wen, Z.~Peng, S.~Huang, S.~Wang, and P.~S. Yu, ``{MISS:} {A} multi-user
  identification network for shared-account session-aware recommendation,'' in
  \emph{DSFAA}, 2021, pp. 228--243.

\bibitem{ma2019pi}
M.~Ma, P.~Ren, Y.~Lin, Z.~Chen, J.~Ma, and M.~d. Rijke, ``$\pi$-net: A parallel
  information-sharing network for shared-account cross-domain sequential
  recommendations,'' in \emph{SIGIR}, 2019, pp. 685--694.

\bibitem{guo_DAGCN_2021}
L.~Guo, L.~Tang, T.~Chen, L.~Zhu, Q.~V.~H. Nguyen, and H.~Yin, ``{DA-GCN:} {A}
  domain-aware attentive graph convolution network for shared-account
  cross-domain sequential recommendation,'' in \emph{IJCAI}, 2021, pp.
  2483--2489.

\bibitem{JMLR_shani05a}
G.~Shani, D.~Heckerman, and R.~I. Brafman, ``An mdp-based recommender system,''
  \emph{JMLR}, pp. 1265--1295, 2005.

\bibitem{chen_playlist_2012}
S.~Chen, J.~L. Moore, D.~Turnbull, and T.~Joachims, ``Playlist prediction via
  metric embedding,'' in \emph{SIGKDD}, 2012, pp. 714--722.

\bibitem{hidasi2016srnn}
B.~Hidasi, A.~Karatzoglou, L.~Baltrunas, and D.~Tikk, ``Session-based
  recommendations with recurrent neural networks,'' in \emph{ICLR}, 2016, pp.
  1--10.

\bibitem{quadrana2017hrnn}
M.~Quadrana, A.~Karatzoglou, B.~Hidasi, and P.~Cremonesi, ``Personalizing
  session-based recommendations with hierarchical recurrent neural networks,''
  in \emph{RecSys}, 2017, pp. 130--137.

\bibitem{li2017neural}
J.~Li, P.~Ren, Z.~Chen, Z.~Ren, T.~Lian, and J.~Ma, ``Neural attentive
  session-based recommendation,'' in \emph{CIKM}, 2017, pp. 1419--1428.

\bibitem{he2018nais}
X.~He, Z.~He, J.~Song, Z.~Liu, Y.~Jiang, and T.~Chua, ``{NAIS:} neural
  attentive item similarity model for recommendation,'' \emph{TKDE}, pp.
  2354--2366, 2018.

\bibitem{Xie2021sequential}
T.~Xie, Y.~XU, L.~Chen, Y.~Liu, and Z.~Zheng, ``Sequential recommendation on
  dynamic heterogeneous information network,'' in \emph{ICDE}, 2021, pp.
  2105--2110.

\bibitem{QiuHLY20}
R.~Qiu, Z.~Huang, J.~Li, and H.~Yin, ``Exploiting cross-session information for
  session-based recommendation with graph neural networks,'' \emph{TOIS}, pp.
  22:1--22:23, 2020.

\bibitem{ChenYCNPL19}
H.~Chen, H.~Yin, T.~Chen, Q.~V.~H. Nguyen, W.~Peng, and X.~Li, ``Exploiting
  centrality information with graph convolutions for network representation
  learning,'' in \emph{ICDE}, 2019, pp. 590--601.

\bibitem{Hsu2021sequential}
C.~Hsu and C.-T. Li, ``Retagnn: Relational temporal attentive graph neural
  networks for holistic sequential recommendation,'' in \emph{WWW}, 2021, pp.
  2968--2979.

\bibitem{wang2014user}
Z.~Wang, Y.~Yang, L.~He, and J.~Gu, ``User identification within a shared
  account: Improving ip-tv recommender performance,'' in \emph{ADBIS}, 2014,
  pp. 219--233.

\bibitem{zhang2012guess}
A.~Zhang, N.~Fawaz, S.~Ioannidis, and A.~Montanari, ``Guess who rated this
  movie: Identifying users through subspace clustering,'' in \emph{UAI}, 2012,
  pp. 944--953.

\bibitem{yang2015adaptive}
Y.~Yang, Q.~Hu, L.~He, M.~Ni, and Z.~Wang, ``Adaptive temporal model for iptv
  recommendation,'' in \emph{WAIM}, 2015, pp. 260--271.

\bibitem{chengzhao2019ppgn}
C.~Zhao, C.~Li, and C.~Fu, ``Cross-domain recommendation via preference
  propagation graphnet,'' in \emph{CIKM}, 2019, pp. 2165--2168.

\bibitem{ChenYS0GM20}
H.~Chen, H.~Yin, X.~Sun, T.~Chen, B.~Gabrys, and K.~Musial, ``Multi-level graph
  convolutional networks for cross-platform anchor link prediction,'' in
  \emph{SIGKDD}, 2020, pp. 1503--1511.

\bibitem{liu2020transfer}
M.~Liu, J.~Li, G.~Li, and P.~Pan, ``Cross domain recommendation via
  bi-directional transfer graph collaborative filtering networks,'' in
  \emph{CIKM}, 2020, pp. 885--894.

\bibitem{lian2017cccfnet}
J.~Lian, F.~Zhang, X.~Xie, and G.~Sun, ``Cccfnet: A content-boosted
  collaborative filtering neural network for cross domain recommender
  systems,'' in \emph{WWW}, 2017, pp. 817--818.

\bibitem{Li2021crossdomain}
S.~Li, L.~Yao, S.~Mu, W.~X. Zhao, Y.~Li, T.~Guo, B.~Ding, and J.~Wen,
  ``Debiasing learning based cross-domain recommendation,'' in \emph{SIGKDD},
  2021, pp. 3190--3199.

\bibitem{LiuSCZ21}
W.~Liu, J.~Su, C.~Chen, and X.~Zheng, ``Leveraging distribution alignment via
  stein path for cross-domain cold-start recommendation,'' in \emph{NeurIPS},
  2021, pp. 19\,223--19\,234.

\bibitem{ZhuWCLZ20}
F.~Zhu, Y.~Wang, C.~Chen, G.~Liu, and X.~Zheng, ``A graphical and attentional
  framework for dual-target cross-domain recommendation,'' in \emph{IJCAI},
  2020, pp. 3001--3008.

\bibitem{zhu2021unified}
F.~Zhu, Y.~Wang, J.~Zhou, C.~Chen, L.~Li, and G.~Liu, ``A unified framework for
  cross-domain and cross-system recommendations,'' \emph{TKDE}, pp. 0--14,
  2021.

\bibitem{Zhang2019HierarchicalRL}
J.~Zhang, B.~Hao, B.~Chen, C.~Li, H.~Chen, and J.~Sun, ``Hierarchical
  reinforcement learning for course recommendation in moocs,'' in \emph{AAAI},
  2019, pp. 435--442.

\bibitem{wu2021reinforcement}
Y.~Wu and I.~O. Craig~Macdonald, ``Partially observable reinforcement learning
  for dialog-based interactive recommendation,'' in \emph{RecSys}, 2021, pp.
  241--251.

\bibitem{zheng2018reinforcement}
G.~Zheng, F.~Zhang, Z.~Zheng, Y.~Xiang, N.~J. Yuan, X.~Xie, and Z.~Li, ``{DRN:}
  {A} deep reinforcement learning framework for news recommendation,'' in
  \emph{WWW}, 2018, pp. 167--176.

\bibitem{Zhang2021reinforcement}
W.~Zhang, H.~Liu, F.~Wang, T.~Xu, H.~Xin, D.~Dou, and H.~Xiong, ``Intelligent
  electric vehicle charging recommendation based on multi-agent reinforcement
  learning,'' in \emph{WWW}, 2021, pp. 1856--1867.

\bibitem{Sun2021reinforcement}
Y.~Sun, F.~Zhuang, H.~Zhu, Q.~He, and H.~Xiong, ``Cost-effective and
  interpretable job skill recommendation with deep reinforcement learning,'' in
  \emph{WWW}, 2021, pp. 3827--3838.

\bibitem{ren2019net}
W.~Sun, M.~Ma, P.~Ren, Y.~Lin, Z.~Chen, Z.~Ren, J.~Ma, and M.~De~Rijke,
  ``Parallel split-join networks for shared account cross-domain sequential
  recommendations,'' \emph{TKDE}, pp. 1--20, 2021.

\bibitem{xia_self_AAAI_2021}
X.~Xia, H.~Yin, J.~Yu, Q.~Wang, L.~Cui, and X.~Zhang, ``Self-supervised
  hypergraph convolutional networks for session-based recommendation,'' in
  \emph{AAAI}, 2021, pp. 4503--4511.

\bibitem{kang2018self}
W.-C. Kang and J.~McAuley, ``Self-attentive sequential recommendation,'' in
  \emph{2018 IEEE International Conference on Data Mining}, 2018, pp. 197--206.

\bibitem{guo2020group}
L.~Guo, H.~Yin, Q.~Wang, B.~Cui, Z.~Huang, and L.~Cui, ``Group recommendation
  with latent voting mechanism,'' in \emph{ICDE}, 2020, pp. 121--132.

\bibitem{duchi_adaptive_2011}
J.~C. Duchi, E.~Hazan, and Y.~Singer, ``Adaptive subgradient methods for online
  learning and stochastic optimization,'' \emph{JMLR}, pp. 2121--2159, 2011.

\bibitem{wang2018reinforcement}
X.~Wang, Y.~Chen, J.~Yang, L.~Wu, Z.~Wu, and X.~Xie, ``A reinforcement learning
  framework for explainable recommendation,'' in \emph{ICDM}, 2018, pp.
  587--596.

\bibitem{williams1992simple}
R.~J. Williams, ``Simple statistical gradient-following algorithms for
  connectionist reinforcement learning,'' \emph{Machine Learning}, pp.
  229--256, 1992.

\bibitem{sutton2000policy}
R.~S. Sutton, D.~A. McAllester, S.~P. Singh, and Y.~Mansour, ``Policy gradient
  methods for reinforcement learning with function approximation,'' in
  \emph{NeuralIPS}, 1999, pp. 1057--1063.

\bibitem{feng2018reinforcement}
J.~Feng, M.~Huang, L.~Zhao, Y.~Yang, and X.~Zhu, ``Reinforcement learning for
  relation classification from noisy data,'' in \emph{AAAI}, 2018, pp.
  5779--5786.

\bibitem{verstrepen2015top}
K.~Verstrepen and B.~Goethals, ``Top-n recommendation for shared accounts,'' in
  \emph{RecSys}, 2015, pp. 59--66.

\bibitem{he2015trirank}
X.~He, T.~Chen, M.-Y. Kan, and X.~Chen, ``Trirank: Review-aware explainable
  recommendation by modeling aspects,'' in \emph{CIKM}, 2015, pp. 1661--1670.

\bibitem{hexiangnan2017ncf}
X.~He, L.~Liao, H.~Zhang, L.~Nie, X.~Hu, and T.-S. Chua, ``Neural collaborative
  filtering,'' in \emph{WWW}, 2017, pp. 173--182.

\bibitem{He2020lightGCN}
X.~He, K.~Deng, X.~Wang, Y.~Li, Y.~Zhang, and M.~Wang, ``Lightgcn: Simplifying
  and powering graph convolution network for recommendation,'' in \emph{SIGIR},
  2020, pp. 639--648.

\bibitem{li2009can}
B.~Li, Q.~Yang, and X.~Xue, ``Can movies and books collaborate? cross-domain
  collaborative filtering for sparsity reduction,'' in \emph{IJCAI}, 2009, pp.
  2052--2057.

\bibitem{yang2017personalized}
S.~Yang, S.~Sarkhel, S.~Mitra, and V.~Swaminathan, ``Personalized video
  recommendations for shared accounts,'' in \emph{MM}, 2017, pp. 256--259.

\bibitem{glorot2010understanding}
X.~Glorot and Y.~Bengio, ``Understanding the difficulty of training deep
  feedforward neural networks,'' in \emph{AISTATS}, 2010, pp. 249--256.

\bibitem{kingma2014adam}
D.~P. Kingma and J.~Ba, ``Adam: {A} method for stochastic optimization,'' in
  \emph{ICLR}, 2015.

\end{thebibliography}

% biography section
% 
% If you have an EPS/PDF photo (graphicx package needed) extra braces are
% needed around the contents of the optional argument to biography to prevent
% the LaTeX parser from getting confused when it sees the complicated
% \includegraphics command within an optional argument. (You could create
% your own custom macro containing the \includegraphics command to make things
% simpler here.)
%\begin{IEEEbiography}[{\includegraphics[width=1in,height=1.25in,clip,keepaspectratio]{mshell}}]{Michael Shell}
% or if you just want to reserve a space for a photo:
\vspace{-1cm}
\begin{IEEEbiography}
[{\includegraphics[width=1in,clip,keepaspectratio]{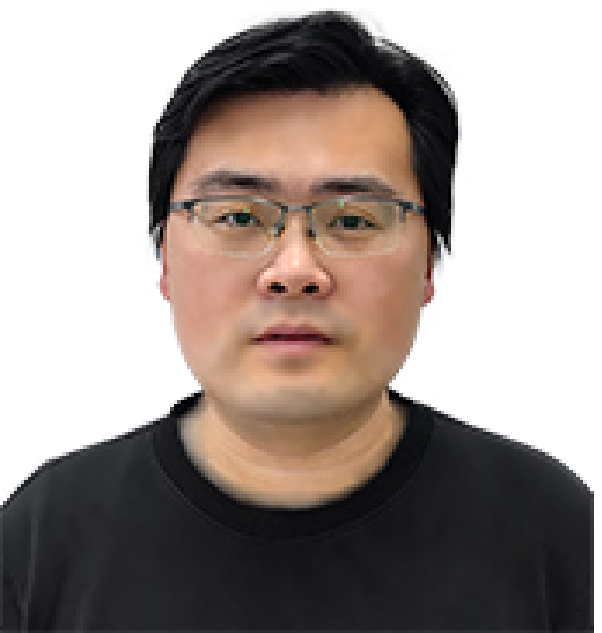}}]
{Lei Guo}
received his Ph.D. degree in computer science from Shandong University, China, in 2015. He is currently an Associate Professor and a Master Supervisor with Shandong Normal University, China. He is a Director of Shandong Artificial Intelligence Society and a member of the Social Media Processing Committee of the Chinese Information Society. His research interests include information retrieval, social networks, and recommender systems.
\vspace{-0.5cm}
\end{IEEEbiography}
%IEEEbiography
% % if you will not have a photo at all:
% \begin{IEEEbiographynophoto}{John Doe}
% Biography text here.
% \end{IEEEbiographynophoto}
\vspace{-1cm}
\begin{IEEEbiography}
[{\includegraphics[width=1in,clip,keepaspectratio]{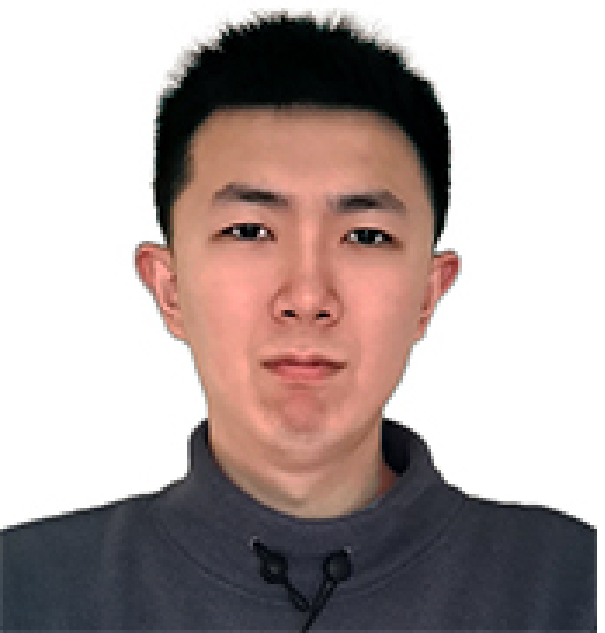}}]
{Jinyu Zhang}
is currently a computer science Master's candidate at the School of information science and Engineering, Shandong Normal University, China. 
His research interests include sequential recommendation and cross-domain recommendation.
\end{IEEEbiography}
\vspace{-1cm}
\begin{IEEEbiography}
[{\includegraphics[width=1in,height=1.25in,clip,keepaspectratio]{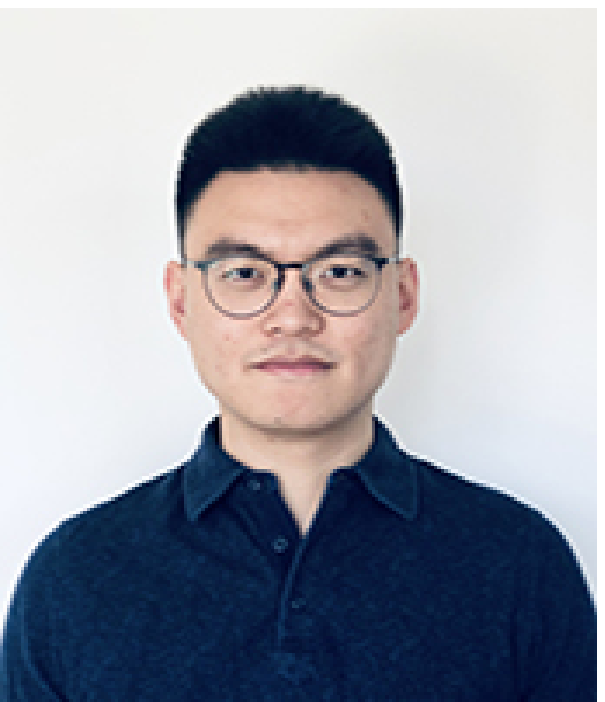}}]
{Tong Chen}
received his Ph.D. degree in computer science from The University of Queensland in 2020.
He is currently a Lecturer with the Data Science research
group, School of Information Technology and Electrical Engineering, The University of Queensland. His research interests include data mining, recommender systems, user behavior modelling and predictive analytics.
\end{IEEEbiography}
\vspace{-1cm}
\begin{IEEEbiography}
[{\includegraphics[width=1in,clip,keepaspectratio]{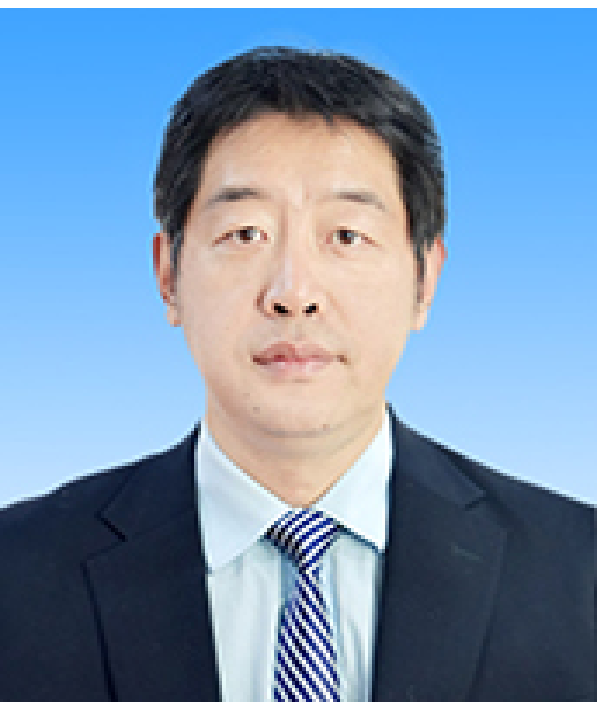}}]
{Xinhua Wang}
received the Ph.D. degree in management science and engineering from Shandong Normal University, China, in 2008. He was a Senior Visiting Scholar with Peking University, from 2008 to 2009. He is currently a Professor and a Master Supervisor with the School of Information Science and Engineering, Shandong Normal University. His research interests include distributed networks and recommender systems.
\end{IEEEbiography}
\vspace{-1cm}
\begin{IEEEbiography}
[{\includegraphics[width=1in,height=1.25in,clip,keepaspectratio]{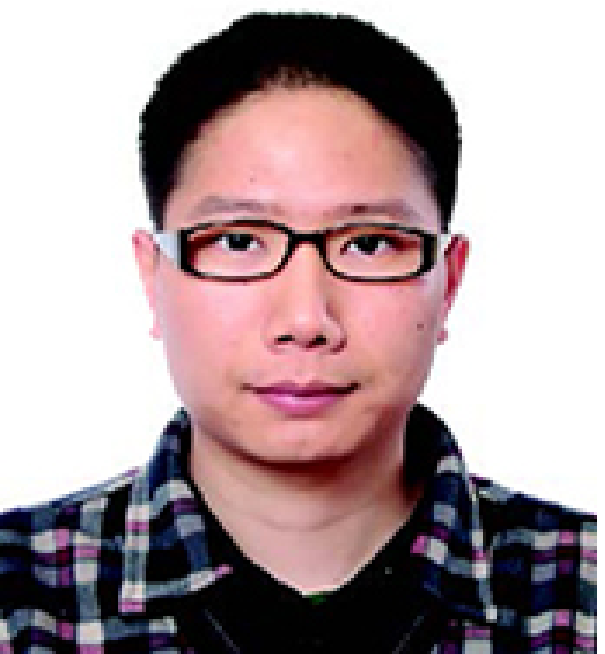}}]
{Hongzhi Yin}
received his Ph.D. degree in computer science from Peking University in 2014. He is an
Associate Professor and Future Fellow with the University
of Queensland. He received the Australian Research Council Future Fellowship and Discovery
Early-Career Researcher Award in 2016 and 2021, respectively. His research interests include recommendation system, user profiling, topic models, deep
learning, social media mining, and location-based services.
\end{IEEEbiography}
% % insert where needed to balance the two columns on the last page with
% % biographies
% %\newpage

% \begin{IEEEbiographynophoto}{Jane Doe}
% Biography text here.
% \end{IEEEbiographynophoto}

% You can push biographies down or up by placing
% a \vfill before or after them. The appropriate
% use of \vfill depends on what kind of text is
% on the last page and whether or not the columns
% are being equalized.

%\vfill

% Can be used to pull up biographies so that the bottom of the last one
% is flush with the other column.
%\enlargethispage{-5in}

% that's all folks
\end{document}